\pdfoutput=1

\documentclass[11pt,a4paper]{article}
\usepackage{jheppub}

\usepackage{url}
\usepackage{bm}

\allowdisplaybreaks[2]

\def\mint{\int_{-\infty}^\infty\!\cdots\!\int_{-\infty}^\infty}

\newcommand{\be}{\begin{equation}}
\newcommand{\ee}{\end{equation}}
\newcommand{\ba}{\begin{aligned}}
\newcommand{\ea}{\end{aligned}}

\def\({\left(}
\def\){\right)}

\newcommand{\del}{\partial}

\DeclareMathOperator{\im}{Im}

\DeclareMathOperator{\eE}{\mathbf{E}}
\DeclareMathOperator{\eK}{\mathbf{K}}
\DeclareMathOperator{\ePi}{\mathbf{\Pi}}

\preprint{RUP-26-7}

\title{Exact WKB and Quantum Periods for Extremal Black Hole Quasinormal Modes}

\author{Yasuyuki Hatsuda}
\author{and Tomohito Shiga}

\affiliation{Department of Physics, Rikkyo University, Toshima, Tokyo 171-8501, Japan}

\emailAdd{yhatsuda@rikkyo.ac.jp, tompaul0530@rikkyo.ac.jp}

\abstract{
We apply exact WKB analysis to the spectral problem arising in black hole perturbation theory.
The boundary conditions for quasinormal modes lead to exact quantization conditions for the complex frequencies.
To solve these conditions, one needs to evaluate the so-called quantum periods, or Voros symbols.
For scalar perturbations of extremal Reissner--Nordstr\"om and Kerr black holes, we compute these quantities up to very high orders in the WKB expansion and perform Borel--Pad\'e resummation.
The resulting resummed quantization conditions successfully reproduce the correct quasinormal mode frequencies with high precision.
}

\begin{document}

\maketitle

\renewcommand{\thefootnote}{\arabic{footnote}}
\setcounter{footnote}{0}
\setcounter{section}{0}

\section{Introduction}
Quasinormal modes (QNMs) are among the most fundamental observables in black hole perturbation theory~\cite{Kokkotas:1999bd, Berti:2009kk, Konoplya:2011qq}. They are defined by imposing purely ingoing behavior at the horizon and purely outgoing behavior at infinity, and their complex frequencies encode both the oscillation and decay of black hole ringdown. While QNM spectra have traditionally been studied by numerical and semi-analytic methods, it is highly desirable to formulate the problem in a controlled analytic framework. Exact WKB analysis~\cite{Voros:1983} offers such a framework. In this approach, the boundary conditions for QNMs naturally lead to exact quantization conditions (EQCs)~\cite{Imaizumi:2022qbi, Imaizumi:2022dgj, Miyachi:2025ptm, Miyachi:2025dyk}, whose basic building blocks are quantum periods, or Voros symbols. 

There are two independent approaches to evaluating the quantum periods. One efficient method is to express them in terms of the Nekrasov partition function~\cite{Nekrasov:2002qd} in the so-called Nekrasov--Shatashvili limit~\cite{Nekrasov:2009rc}, or equivalently in terms of the semiclassical (irregular) conformal block~\cite{Zamolodchikov:1986, Zamolodchikov:1987}, through the correspondence with Seiberg--Witten (SW) theory~\cite{Seiberg:1994rs, Seiberg:1994aj} and two-dimensional CFT~\cite{Belavin:1984vu}. This approach is useful because we know a combinatorial formula of the Nekrasov partition function or a recursive method to obtain the conformal block. 
In the black hole QNM problem, this approach works well in practice~\cite{Aminov:2020yma}.
However, as emphasized in~\cite{Arnaudo:2025kof}, the instanton expansion parameter of the Nekrasov partition function is not always a suitable expansion parameter for the black hole problem. In particular, the convergence deteriorates as the overtone number increases. To circumvent this difficulty, one may instead re-expand the partition function in terms of a different parameter, but no combinatorial closed formula is known in such a case.

The other approach is to evaluate the cycle integrals defining the quantum periods directly.
For genus-one spectral curves, these quantum periods can in principle be expressed in terms of elliptic integrals.
In practice, however, obtaining explicit analytic expressions is not always straightforward, since the relevant integrals typically involve complicated parameter dependence.
Nevertheless, in certain special cases the structure of the spectral curve simplifies sufficiently, and the quantum periods can then be computed in a much more systematic and explicit manner~\cite{Grassi:2019coc}. This direct approach ultimately leads to TBA equations for the Borel resummed quantum periods~\cite{Ito:2018eon, Ito:2019jio}.

In this work, we focus on the QNM problem for scalar perturbations of four-dimensional asymptotically flat extremal black holes.\footnote{Very recently, the QNM problem for scalar perturbations of the extremal Reissner--Nordstr\"om black hole was analyzed using its connection with $\mathcal{N}=2$ SW theory~\cite{Wang:2026kue}. In that work, the Nekrasov--Shatashvili free energy was used to solve the problem. Although there is a slight overlap, the approach is clearly different from ours.} The advantage of the extremal limit is that the computation of the quantum periods becomes much simpler. We compute the quantum periods up to very high orders in the WKB expansion. Such high-order data allow us to perform Borel--Pad\'e resummation of the quantum period. As a result, we can solve the resummed quantization conditions, and they provide the correct QNM frequencies with high precision. %We also discuss how to extend this approach to non-extremal cases, as well as to gravitational and electromagnetic perturbations.

The rest of this paper is organized as follows.
In Section~2, we study scalar perturbations of the extremal Reissner--Nordstr\"om black hole.
We first reduce the radial equation to the doubly confluent Heun equation, then review the basic ingredients of exact WKB analysis, derive the exact quantization conditions for the QNM frequencies, and finally evaluate the quantum periods and compare the resulting resummed quantization conditions with numerical QNM data.
In Section~3, we carry out the analogous analysis for the extremal Kerr black hole.
Section~4 is devoted to our conclusions.
In Appendix~\ref{app:ACQP}, we explain how to compute the quantum periods analytically.

\section{Extremal Reissner--Nordstr\"om Black Hole}

In this section, we analyze scalar perturbations of the extremal Reissner--Nordstr\"om black hole using exact WKB analysis. We first reduce the radial equation to the doubly confluent Heun equation and review the basic ingredients of exact WKB analysis relevant to our problem. We then derive the exact quantization condition for the QNM frequencies and evaluate the corresponding quantum periods. Finally, we compare the resulting Borel--Pad\'e resummed quantization condition with numerical QNM data.

\subsection{Reduction to Doubly Confluent Heun Equation}

We begin with the extremal Reissner--Nordstr\"om background,
\begin{align}
ds^2=-f(r)dt^2+\frac{dr^2}{f(r)}+r^2 (d\theta^2+\sin^2 \theta d\phi^2),\qquad
f(r)=1-\frac{2M}{r}+\frac{Q^2}{r^2},
\end{align}
where $M$ and $Q$ are the mass and the charge of the black hole.
At linear order in the perturbation, the scalar field equation is simply given by the Klein--Gordon equation on this background:
\begin{align}
(-g)^{-1/2}\del_\mu [ (-g)^{1/2}g^{\mu\nu}\del_\nu ]\Phi=0.
\label{eq:KG}
\end{align}
By the separation of variables
\begin{align}
\Phi=e^{-i\omega t} \frac{\phi(r)}{r} Y_{\ell}^{m}(\theta, \phi),
\end{align}
where $Y_{\ell}^{m}(\theta, \phi)$ is a spherical harmonic,
we obtain the radial ordinary differential equation (ODE),
\begin{align}
\biggl[ f(r) \frac{d}{dr} \biggl( f(r) \frac{d}{dr} \biggr)+\omega^2 - V_\text{RN}(r) \biggr] \phi(r)=0,
\label{eq:radial-ODE}
\end{align}
where
\begin{equation}
\begin{aligned}
V_\text{RN}(r)&=f(r)\biggl( \frac{\ell(\ell+1)}{r^2}+\frac{f'(r)}{r}\biggr) \\
&=\biggl( 1-\frac{2M}{r}+\frac{Q^2}{r^2} \biggr)\biggl( \frac{\ell(\ell+1)}{r^2}+\frac{2M}{r^3}-\frac{2Q^2}{r^4}\biggr).
\end{aligned}
\label{eq:V_RN}
\end{equation}
For $0 \leq Q <M$, $f(r)=0$ has two distinct real roots,
\begin{align}
r_\pm=M \pm \sqrt{M^2-Q^2},
\end{align}
where $r=r_+$ corresponds to the event horizon.
We also introduce the tortoise coordinate $r_*$ by
\begin{align}
r_*=r+\frac{1}{r_+-r_-}\Bigl(r_+^2 \log(r-r_+)-r_-^2 \log(r-r_-) \Bigr).
\end{align}
The limit $r_* \to -\infty$ corresponds to the event horizon, while $r_* \to +\infty$ corresponds to spatial infinity.
Then, the QNM boundary conditions are given by
\begin{align}
\phi(r) \to \begin{cases} e^{-i\omega r_*} \quad &(r_* \to -\infty) \\
e^{i\omega r_*} \quad &(r_* \to +\infty) \end{cases}.
\label{eq:QNM-1}
\end{align}
For $0 \leq Q <M$, the ODE \eqref{eq:radial-ODE} turns out to have two regular singular points at $r=r_{\pm}$ and one irregular singular point at $r=\infty$. Therefore it is equivalent to the confluent Heun equation.

In the extremal case $Q=M$, the situation changes drastically. In this case, $r_-=r_+$, $f(r)$ has a double zero at $r=M$, and the tortoise coordinate is given by
\begin{align}
r_*=r-\frac{M^2}{r-M}+2M \log(r-M)+\text{const}.
\end{align}
The ODE \eqref{eq:radial-ODE} has two irregular singular points at $r=M, \infty$. It is equivalent to the doubly confluent Heun equation (DCHE).
To see it more explicitly, let us change the variables,
\begin{align}
r=M(1+z),\qquad \phi(r)=\biggl(1+\frac{1}{z}\biggr) \psi(z).
\end{align}
We immediately obtain the following ODE:
\begin{align}
\biggl( -\frac{d^2}{dz^2}-(M\omega)^2 \biggl(1+\frac{1}{z} \biggr)^4+\frac{\ell(\ell+1)}{z^2} \biggr) \psi(z)=0
\label{eq:RN-scalar}
\end{align}
For this ODE, one can easily see that $z=0,\infty$ are irregular singular points.
Near $z=0$, we have the two asymptotic solutions,
\begin{align}
\psi(z) \sim e^{\pm iM\omega/z} z^{1\mp 2iM\omega} \quad (z \to 0). 
\label{eq:sol-near-0}
\end{align}
Similarly, the two asymptotic solutions around $z=\infty$ are given by
\begin{align}
\psi(z) \sim e^{\pm iM\omega z} z^{\pm 2iM\omega} \quad (z \to \infty).
\label{eq:sol-near-inf}
\end{align}
By carefully tracking the change of variables, one finds that the QNM boundary conditions \eqref{eq:QNM-1} are mapped to the following condition in the $z$-variable:
\begin{align}
\psi(z) \to \begin{cases} e^{iM\omega/z} z^{1-2iM\omega} \quad &(z \to 0^+) \\
e^{iM\omega z} z^{2iM\omega} \quad &(z \to +\infty) \end{cases},
\label{eq:QNM-2}
\end{align}
where $z \to 0^+$ means that $z$ goes to zero along the positive real axis.
%We will write down an exact equation to satisfy this boundary condition by using the exact WKB connection formula.

\subsection{Review of Exact WKB}
We would like to apply exact WKB analysis to the perturbation equation \eqref{eq:RN-scalar} to solve the QNM problem.
In this subsection, we review the basics of exact WKB analysis~\cite{Voros:1983, Aoki:1991, Delabaere:1997srq}.

Let us consider the one-dimensional Schr\"odinger-type equation:
\begin{align}
\biggl(-\hbar^2 \frac{d^2}{dx^2}+Q(x) \biggr) \psi(x)=0,
\label{eq:Sch}
\end{align}
where we assume that $Q(x)$ takes the following form:
\begin{align}
Q(x)=Q_0(x)+\sum_{k=1}^\infty \hbar^{2k} Q_{2k}(x).
\end{align}
We construct the WKB solution of the form
\begin{align}
\psi(x)=\exp \biggl( \int^x S(x')dx' \biggl),
\end{align}
where $S(x)$ satisfies the Riccati equation,
\begin{align}
\hbar^2 \bigl( S(x)^2+S'(x) \bigr)=Q(x)
\end{align}
We can construct a formal power series solution,
\begin{align}
S(x)=\sum_{j=0}^\infty \hbar^{j-1} S_{j-1}(x).
\end{align}
At the leading order, we find
\begin{align}
S_{-1}(x)=\pm \sqrt{Q_0(x)},
\end{align}
which provides two independent WKB solutions.
We divide $S(x)$ into the odd/even power parts,
\begin{align}
S_\text{odd}(x)=\sum_{k=0}^\infty \hbar^{2k-1}S_{2k-1}(x),\qquad
S_\text{even}(x)=\sum_{k=0}^\infty \hbar^{2k}S_{2k}(x).
\label{eq:S_odd/even}
\end{align}
Then it is well-known that the following identity holds:
\begin{align}
S_\text{even}(x)=-\frac{1}{2} \frac{d}{dx} \log S_\text{odd}(x).
\end{align}
We obtain the following two independent WKB solutions:
\begin{align}
\psi_\pm (x;x_0)=\frac{1}{\sqrt{S_\text{odd}(x)}} \exp \biggl( \pm \int_{x_0}^x S_\text{odd}(x')dx'\biggr),
\label{eq:WKB-sol}
\end{align}
where $x_0$ is a reference point.

The WKB solutions have formal series in $\hbar$,
\begin{equation}
\begin{aligned}
\psi_{\pm}(x;x_0)=\exp (\pm \hbar^{-1}y_0(x) ) \sum_{n=0}^\infty \psi_{\pm, n}(x)\hbar^{n+\frac{1}{2}},
\end{aligned}
\label{eq:WKB-series}
\end{equation}
where
\begin{align}
y_0(x):=\int_{x_0}^x \sqrt{Q_0(x')}dx'.
\end{align}
These series turn out to be divergent in general. We use Borel resummation to construct analytic solutions.
The Borel resummations of \eqref{eq:WKB-series} are given by
\begin{align}
\Psi_\pm(x;x_0)=\mathcal{S}[\psi_\pm(x;x_0)]=\int_{\mp y_0(x)}^\infty e^{-\frac{y}{\hbar}} \widehat{\psi}_{\pm}(x,y)dy,
\end{align}
where
\begin{align}
\widehat{\psi}_{\pm}(x,y):=\sum_{n=0}^\infty \frac{\psi_{\pm, n}(x)}{\Gamma(n+\frac{1}{2})} (y \pm y_0(x))^{n-\frac{1}{2}}
\end{align}
are referred to as the Borel transforms. The symbol $\mathcal{S}$ represents the Borel resummation.
$\Psi_\pm(x;x_0)$ are analytic solutions to the original equation \eqref{eq:Sch}.

An important point is that the Borel resummed WKB solutions $\Psi_\pm(x;x_0)$ exhibit the Stokes phenomenon.
The zeros of $Q_0(x)$ are called turning points. In particular, a simple zero $x=a$ of $Q_0(x)$ is called a simple turning point.
Now we define the Stokes curves. The Stokes curves emanating from a turning point $x=a$ are defined by
\begin{align}
\im \biggl[ \int_a^x \sqrt{Q_0(x')} \, dx' \biggr] = 0.
\end{align}
If $x=a$ is a simple turning point, then three Stokes curves emanate from $x=a$. 
On a Stokes curve, the dominant WKB solution is not Borel summable. 
When one crosses a Stokes curve, this solution changes discontinuously. 
This phenomenon is called the Stokes phenomenon. 
Let us denote the dominant and subdominant solutions by $\Psi_{\mathrm{dom}}$ and $\Psi_{\mathrm{sub}}$, respectively. 
\begin{figure}[tb]
\centering
 \includegraphics[width=0.35\linewidth]{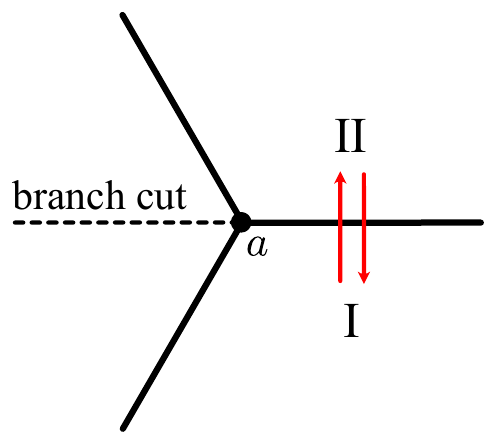}
\caption{Three Stokes curves emanate from a simple turning point. Across each Stokes curve, one of the Borel resummed WKB solutions undergoes a discontinuous jump, which is the Stokes phenomenon.}
\label{fig:connection}
\end{figure}
Then, as shown in Fig.~\ref{fig:connection}, the analytic continuation from sector I to sector II, obtained by crossing the Stokes curve counterclockwise around the turning point, is described by the following connection formula:\footnote{Here we assume that no Stokes curve emanating from a turning point terminates at another turning point.}
\begin{align}
\begin{pmatrix}
\Psi_\mathrm{dom}^\mathrm{I}(x;a) \\
\Psi_\mathrm{sub}^\mathrm{I}(x;a)
\end{pmatrix}
=
\begin{pmatrix}
1 & i \\
0 & 1
\end{pmatrix}
\begin{pmatrix}
\Psi_\mathrm{dom}^\mathrm{II}(x;a) \\
\Psi_\mathrm{sub}^\mathrm{II}(x;a)
\end{pmatrix}.
\end{align}
Conversely, the analytic continuation from sector II to sector I, obtained by crossing the Stokes curve clockwise around the turning point, is given by the inverse relation
\begin{align}
\begin{pmatrix}
\Psi_\mathrm{dom}^\mathrm{II}(x;a) \\
\Psi_\mathrm{sub}^\mathrm{II}(x;a) 
\end{pmatrix}
=
\begin{pmatrix}
1 & -i \\
0 & 1
\end{pmatrix}
\begin{pmatrix}
\Psi_\mathrm{dom}^\mathrm{I}(x;a) \\
\Psi_\mathrm{sub}^\mathrm{I}(x;a)
\end{pmatrix}.
\end{align}
Note that $\Psi_{\mathrm{dom}}$ and $\Psi_{\mathrm{sub}}$ should not be confused with $\Psi_{+}$ and $\Psi_{-}$. The latter are simply the solutions constructed from the definition \eqref{eq:WKB-sol}. In applying the connection formula, one must always keep track of which of $\Psi_{\pm}$ is dominant and which is subdominant.

\subsection{Exact Quantization Conditions for QNM Frequencies}
In this subsection, we derive the EQCs that determine the QNM frequencies for the extremal RN black holes. %The perturbation equation is the DCHE \eqref{eq:RN-scalar}. 
To apply exact WKB to this system, we need to introduce a formal Planck parameter by hand. There are infinitely many possibilities to do so.
To fix one natural choice, we use the connection with the quantum Seiberg--Witten curve~\cite{Ashok:2016yxz, Grassi:2021wpw}. The corresponding gauge theory is $\mathcal{N}=2$ $SU(2)$ SQCD with $N_f=2$ flavors. The SW curve of this theory is given by
\begin{equation}
\begin{aligned}
p^2+2\Lambda^2 \cosh 2x+2\Lambda (m_1 e^{x}+m_2 e^{-x})=2u,
\end{aligned}
\end{equation}
where $m_1$ and $m_2$ are flavor masses, $\Lambda$ is the dynamical scale, and $u$ is the Coulomb modulus.
The SW one-form is $\lambda_\text{SW}=pdx$, and we can regard $x$ and $p$ as canonical coordinates. 
The quantization of this SW curve thus leads to the Schr\"odinger equation:
\begin{equation}
\begin{aligned}
\biggl( -\hbar^2 \frac{d^2}{dx^2}+2\Lambda^2 \cosh 2x+2\Lambda (m_1 e^{x}+m_2 e^{-x}) \biggr) \varphi(x)=2u \varphi(x)
\end{aligned}
\end{equation}
We perform the following change of variables:
\begin{equation}
\begin{aligned}
z=e^x,\qquad \varphi(x)=z^{-1/2}\psi(z)
\end{aligned}
\end{equation}
Then we find
\begin{align}
\biggl(-\hbar^2 \frac{d^2}{dz^2}+Q_0(z)+\hbar^2 Q_2(z) \biggr) \psi(z)=0,
\label{eq:Sch-2}
\end{align}
where
\begin{align}
Q_0(z)=\frac{\Lambda^2}{z^4}+\frac{2m_2\Lambda }{z^3}-\frac{2u}{z^2}+\frac{2 m_1\Lambda}{z}+\Lambda^2,\qquad
Q_2(z)=-\frac{1}{4z^2}.
\label{eq:Q0Q2}
\end{align}
Now we compare this result with the perturbation equation \eqref{eq:RN-scalar}.
It turns out that if we take the following parameter identification:
\begin{equation}
\begin{aligned}
\hbar&=1,&\qquad 2u&=6(M\omega)^2-\biggl(\ell+\frac{1}{2}\biggr)^2, \\
\Lambda&=-iM\omega,&\qquad m_1&=m_2=-2iM\omega,\qquad
\end{aligned}
\label{eq:id-1}
\end{equation}
then the equation \eqref{eq:Sch-2} with \eqref{eq:Q0Q2} exactly matches \eqref{eq:RN-scalar}.
Under this identification, the classical function $Q_0(z)$ becomes
\begin{equation}
\begin{aligned}
Q_0(z)=-(M\omega)^2 \biggl(1+\frac{1}{z}\biggr)^4+\frac{(\ell+1/2)^2}{z^2}.
\end{aligned}
\label{eq:Q0-eRN}
\end{equation}
The replacement $\ell(\ell+1) \to (\ell+1/2)^2$ is well known as the Langer prescription~\cite{Langer:1937} in the conventional WKB approximation.

Now we apply exact WKB to \eqref{eq:Sch-2} with \eqref{eq:Q0-eRN} and $Q_2(z)=-1/(4z^2)$. %We keep $\hbar$ as the formal parameter.
%For later convenience, we introduce new variables by
%\begin{equation}
%\begin{aligned}
%m_1=m_2=\mu, \qquad E=2u+2\Lambda^2+\mu^2=-\biggl(\ell+\frac{1}{2}\biggr)^2.
%\end{aligned}
%\end{equation}
We first examine the behavior of the WKB solutions as $z\to0$ and $z\to\infty$.
We fix the branch of $\sqrt{Q_0(z)}$ along the positive real axis. 
As $z\to+\infty$ along this axis, we have
\[
Q_0(z)
\sim
-(M\omega)^2\left(1+\frac{4}{z}+O(z^{-2})\right).
\]
We choose the branch such that
\begin{equation}
\sqrt{Q_0(z)}
=
iM\omega\biggl(1+\frac{2}{z}+O(z^{-2})\biggr),
\qquad z\to+\infty.
\end{equation}
Using the leading WKB form
\begin{equation}
\psi_\pm
\sim
Q_0(z)^{-1/4}
\exp\left(
\pm \int^z \sqrt{Q_0(z')}\,dz'
\right),
\end{equation}
we find
\begin{equation}
\begin{aligned}
\psi_\pm 
&\sim \exp\biggl[ \pm iM\omega \int^z \biggl( 1+\frac{2}{z'} \biggr)dz' \biggr]  \\
&\sim z^{\pm 2i M\omega} \exp (\pm iM\omega z),
\end{aligned}
\end{equation}
which precisely agrees with \eqref{eq:sol-near-inf}.

We then analytically continue the same branch to the near-horizon region $z\to0$ without crossing branch cuts. Near $z=0$, we have
\begin{equation}
\begin{aligned}
Q_0(z)
&\sim
-(M\omega)^2\biggl(\frac{1}{z^4}+\frac{4}{z^3}+O(z^{-2}) \biggr),\\
\sqrt{Q_0(z)}
&\sim
i M\omega \biggl(\frac{1}{z^2}+\frac{2}{z}+O(1) \biggr).
\end{aligned}
\end{equation}
Since $Q_0(z)^{-1/4}\sim z$ in this limit, the leading WKB solutions behave as
\begin{equation}
\begin{aligned}
\psi_\pm 
&\sim z \exp\biggl[\pm iM\omega \int^z \biggl( \frac{1}{z'^2}+\frac{2}{z'} \biggr) dz' \biggr] \\
&\sim z^{1\pm 2iM\omega} \exp \biggl( \mp \frac{iM\omega}{z} \biggr).
\end{aligned}
\end{equation}
This behavior precisely agrees with \eqref{eq:sol-near-0} for the analytic solutions.

\begin{figure}[tb]
  \begin{minipage}[b]{0.45\linewidth}
    \centering
    \includegraphics[width=0.95\linewidth]{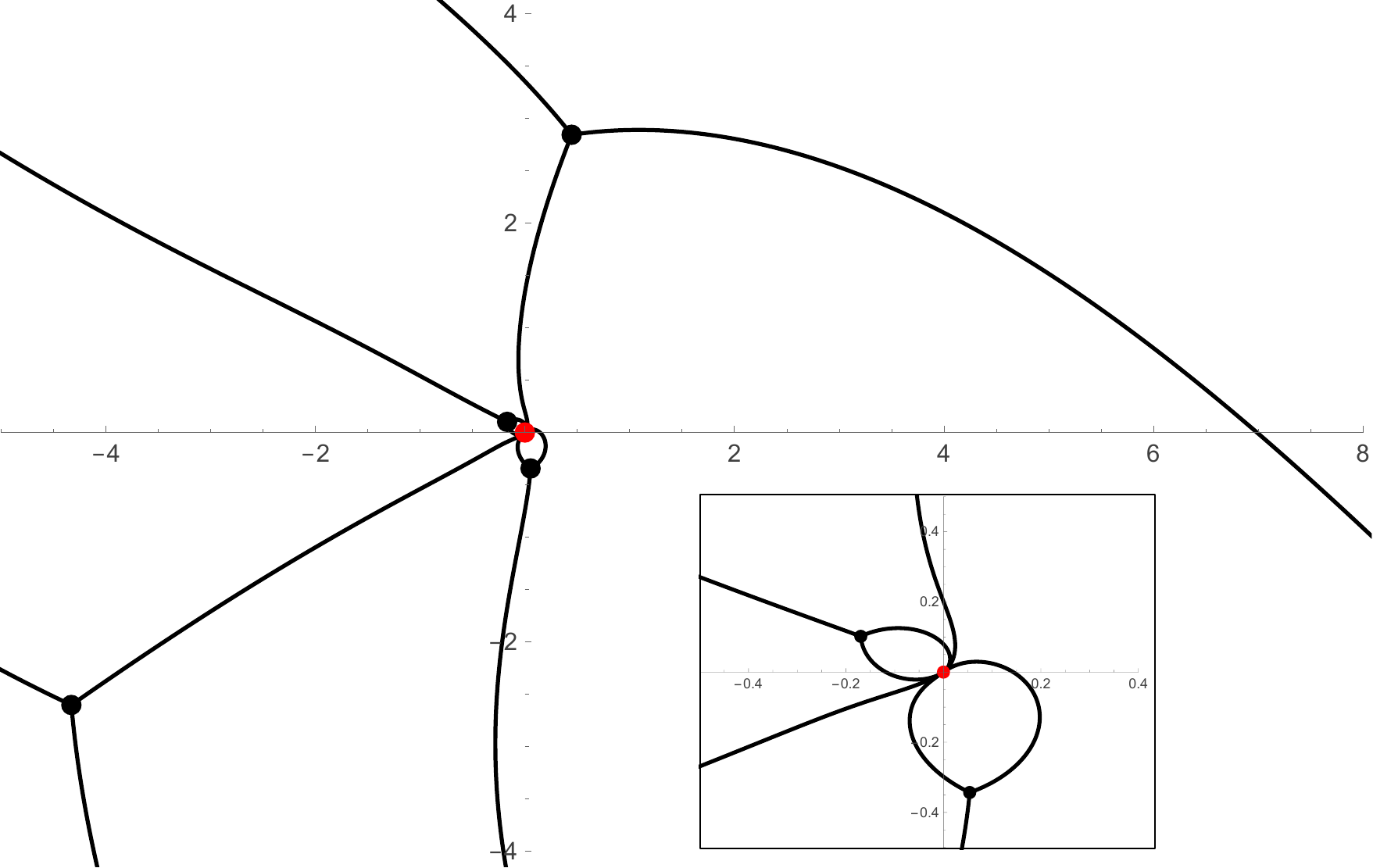}\\
    {\footnotesize (a) $\ell=0$ and $M\omega=0.1-0.1i$.}
  \end{minipage}
    \begin{minipage}[b]{0.45\linewidth}
    \centering
    \includegraphics[width=0.95\linewidth]{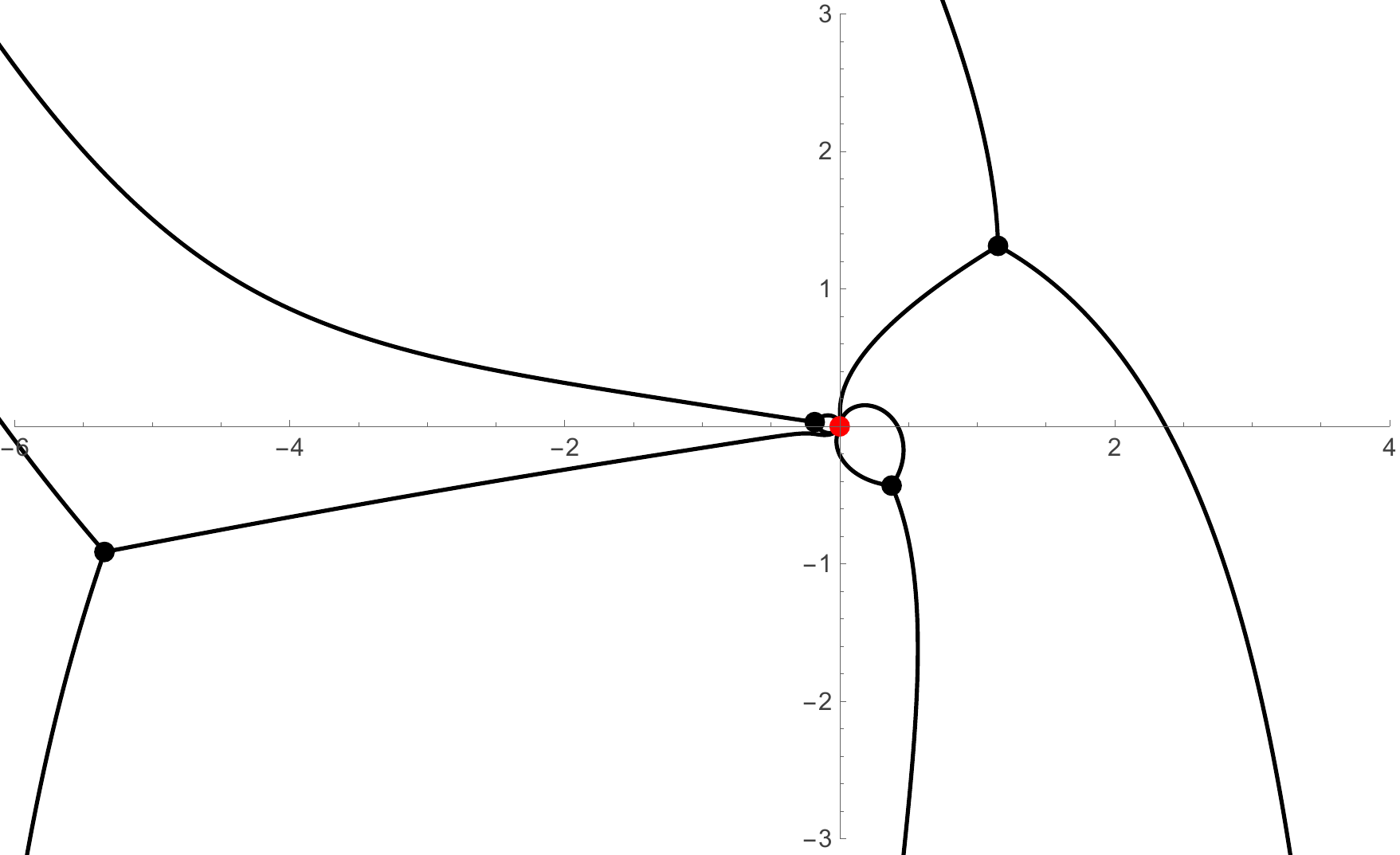}\\
    {\footnotesize (b) $\ell=1$ and $M\omega=0.4-0.1i$.}
  \end{minipage}\\
  \begin{minipage}[b]{0.45\linewidth}
    \centering
    \includegraphics[width=0.95\linewidth]{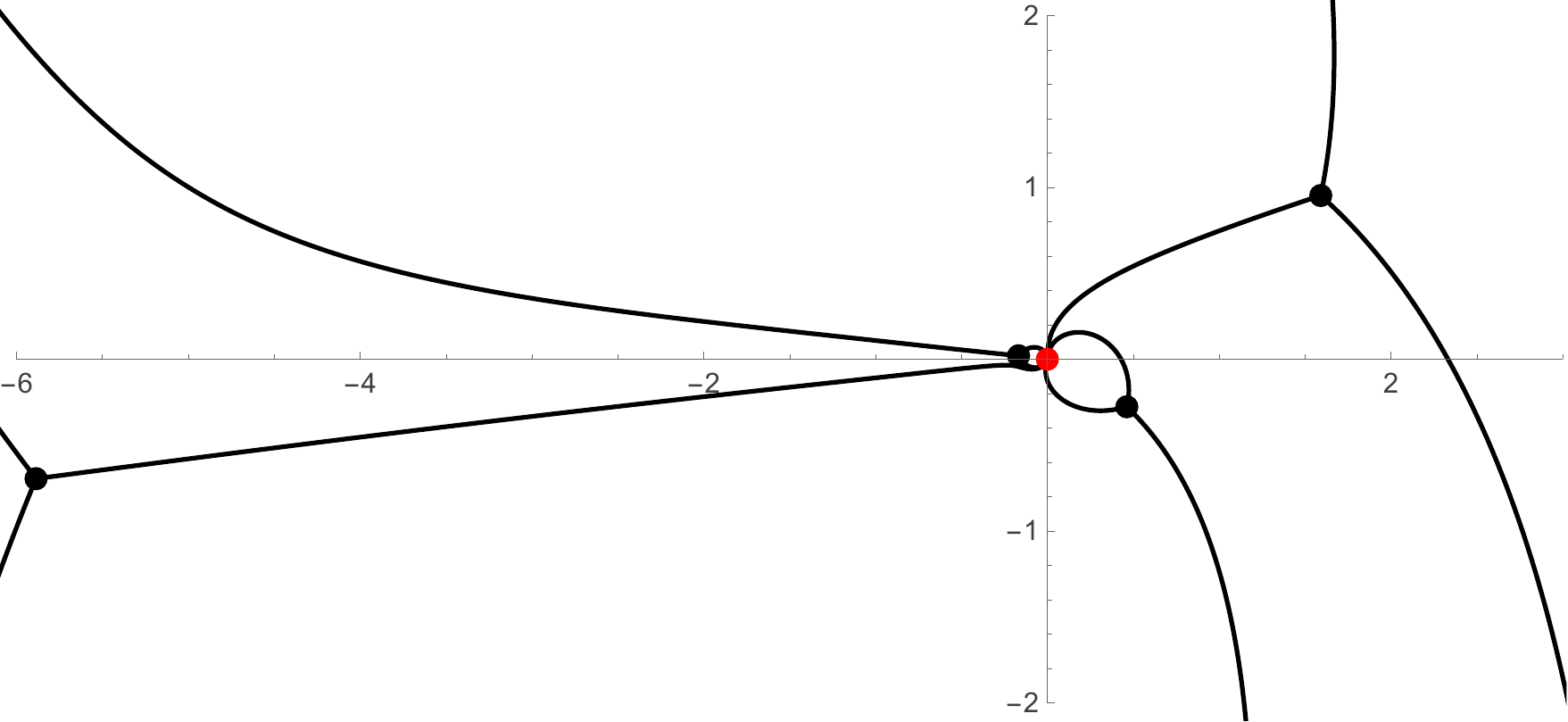}\\
    {\footnotesize (c) $\ell=2$ and $M\omega=0.6-0.1i$.}
  \end{minipage}
    \begin{minipage}[b]{0.45\linewidth}
    \centering
    \includegraphics[width=0.95\linewidth]{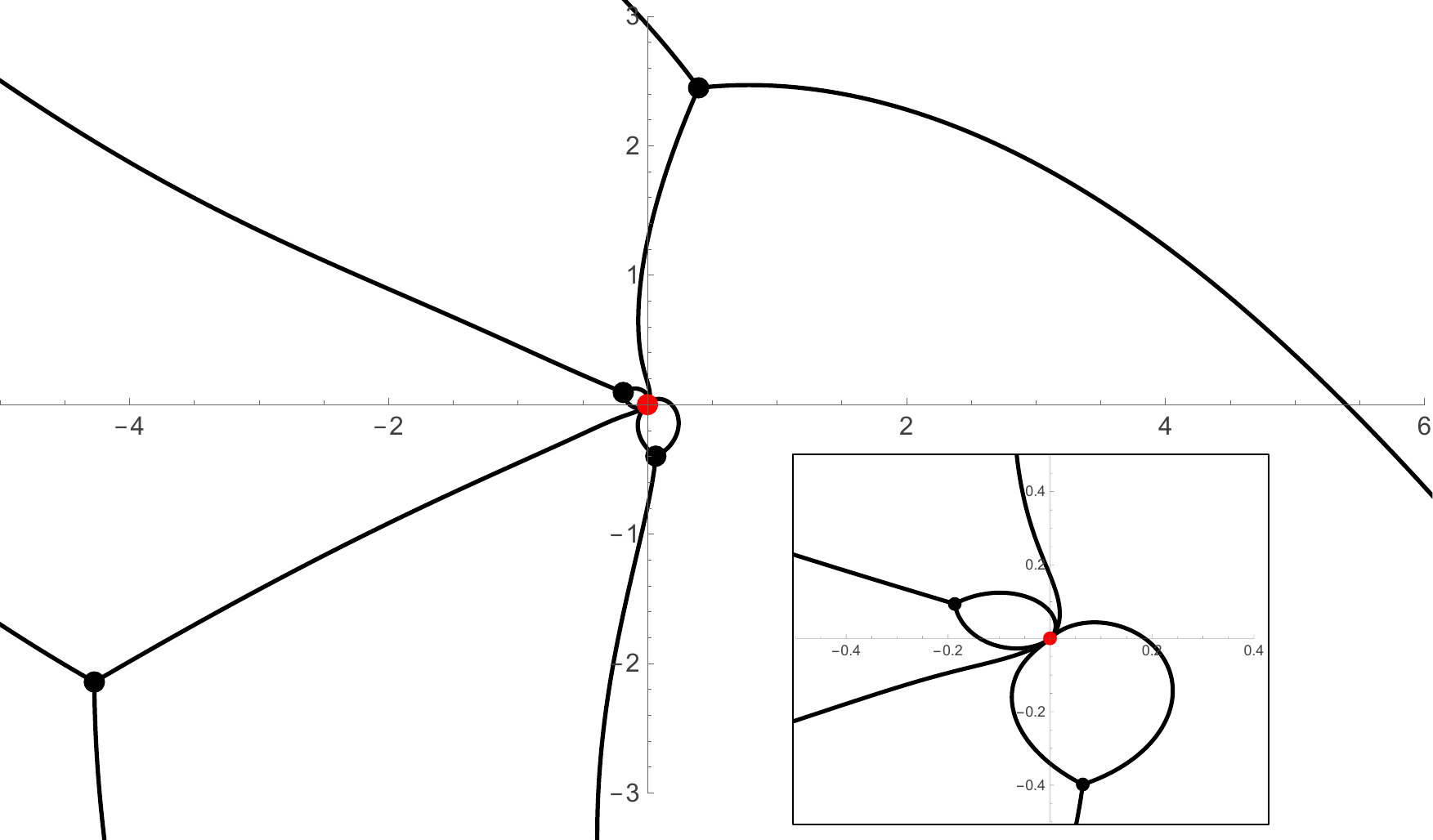}\\
    {\footnotesize (d) $\ell=2$ and $M\omega=0.6-0.5i$.}
  \end{minipage}\\
  \begin{minipage}[b]{0.45\linewidth}
    \centering
    \includegraphics[width=0.95\linewidth]{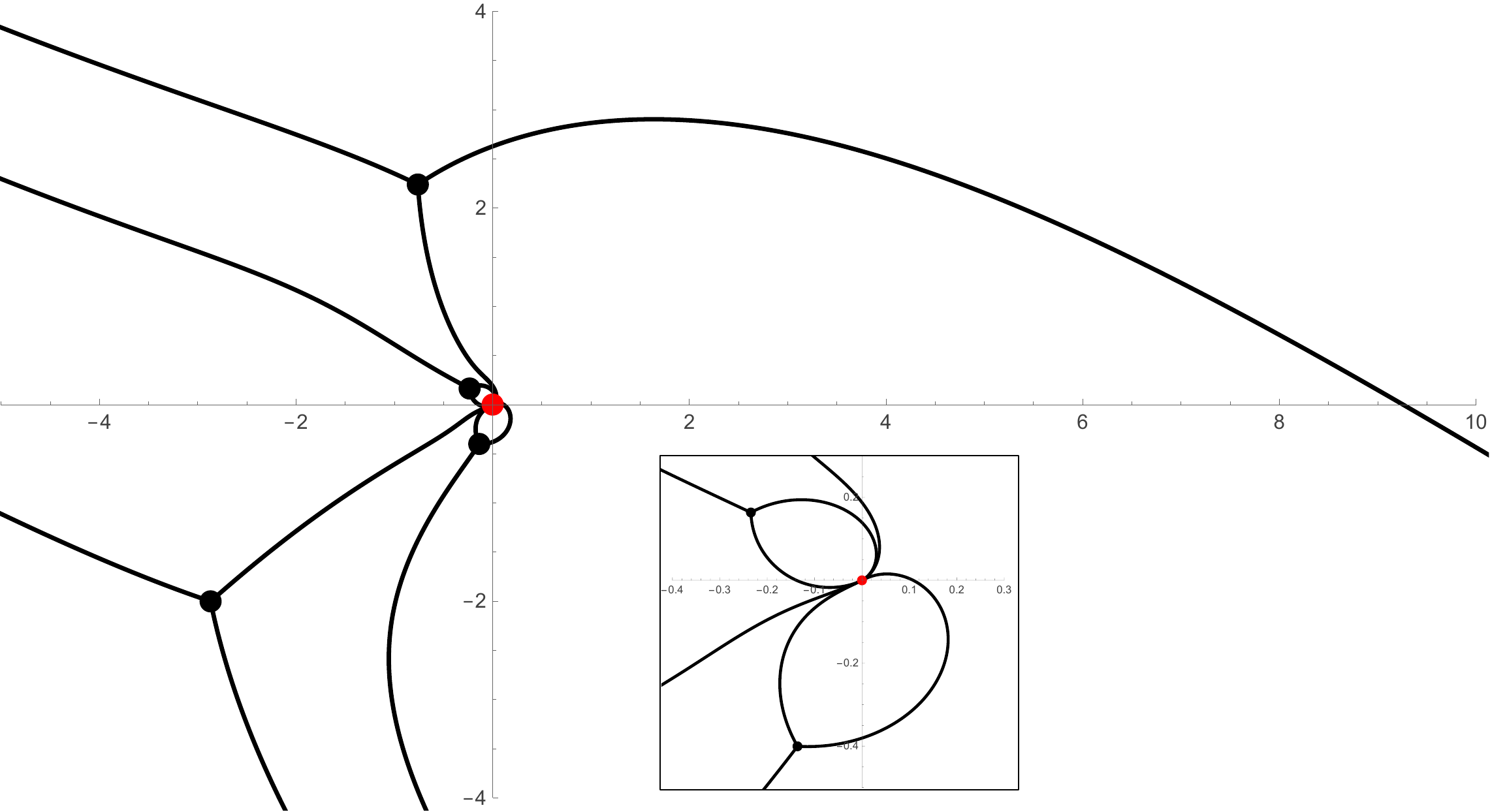}\\
    {\footnotesize (e) $\ell=2$ and $M\omega=0.6-i$.}
  \end{minipage}
    \begin{minipage}[b]{0.45\linewidth}
    \centering
    \includegraphics[width=0.95\linewidth]{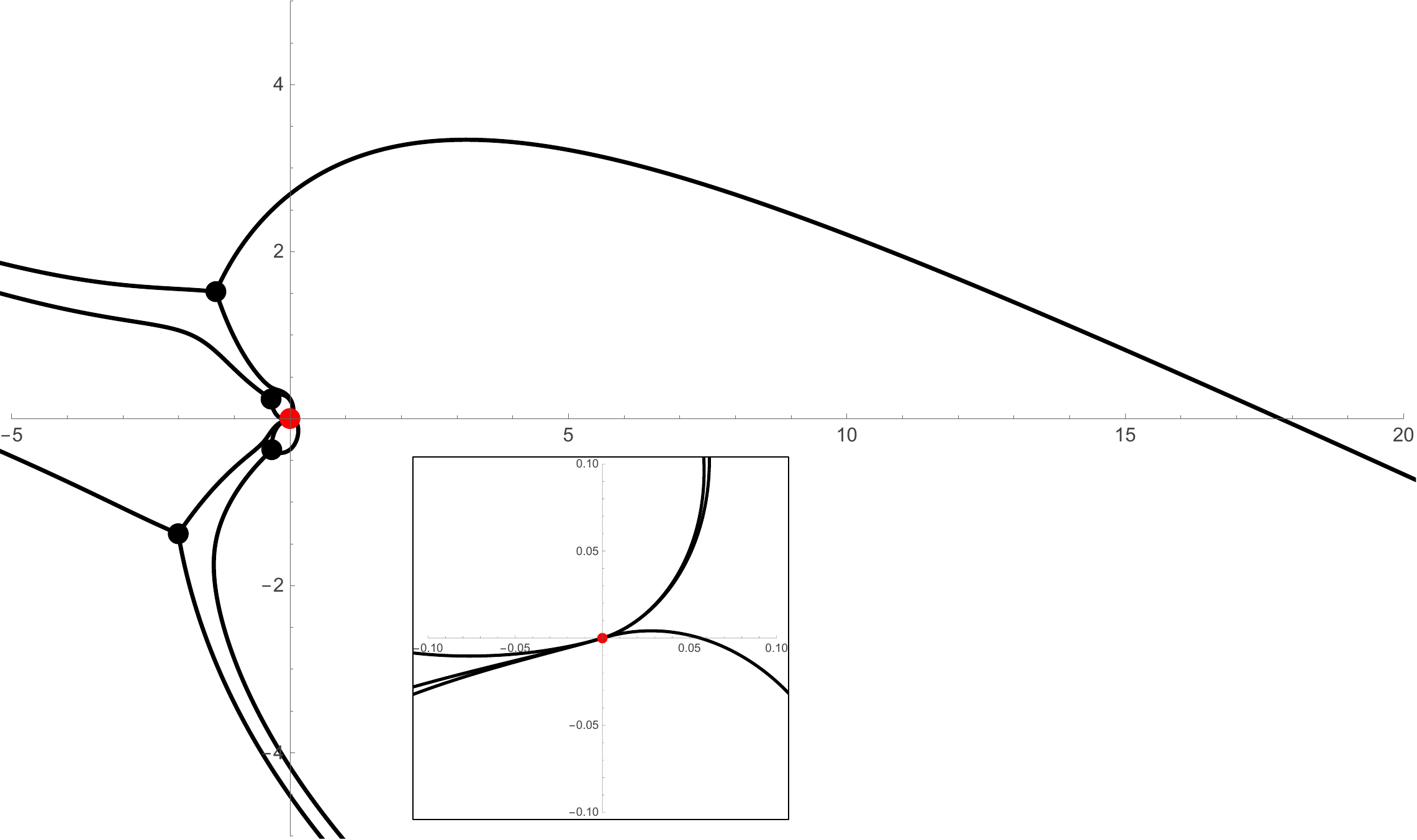}\\
    {\footnotesize (f) $\ell=2$ and $M\omega=0.6-2i$.}
  \end{minipage}\\
  \caption{Stokes graphs of the extremal RN black hole for selected values of $\ell$ and $M\omega$.}
  \label{fig:Stokes-RN}
\end{figure}

From these results, we conclude that the QNM boundary condition \eqref{eq:QNM-2} is translated into the WKB solutions as
\begin{align}
\psi(z) \to \begin{cases} \Psi_-(z) \quad &(z \to 0^+) \\
\Psi_+(z) \quad &(z \to +\infty) \end{cases},
\label{eq:QNM-3}
\end{align}
where we have omitted the reference point, which is not important for the boundary condition.
We analytically continue the Borel resummed WKB solutions from $z=0$ to $z=\infty$ along the positive real axis.
For this purpose, we have to examine the Stokes graphs for given parameters.
It is easy to see that $Q_0(z)$ has four turning points in general. These are given by
\begin{equation}
\begin{aligned}
\alpha_\pm=\frac{-A-2\pm \sqrt{A(A+4)}}{2},\qquad
\beta_\pm=\frac{A-2 \pm \sqrt{A(A-4)}}{2},
\end{aligned}
\end{equation}
where
\begin{equation}
\begin{aligned}
A:=\frac{\ell+1/2}{M\omega}.
\end{aligned}
\end{equation}
%Throughout this paper, we assume $A \ne 0, \pm4$.
Note that we have the symmetric relations $\alpha_-=\alpha_+^{-1}$ and $\beta_-=\beta_+^{-1}$.
In Fig.~\ref{fig:Stokes-RN}, we show Stokes graphs for various values of $\ell$ and $M\omega$.
We find that the topology of the Stokes graphs remains unchanged as $\ell$ and $M\omega$ are varied. 
Although we do not have a rigorous proof of this fact, we assume it.

\begin{figure}[tb]
\centering
 \includegraphics[width=0.9\linewidth]{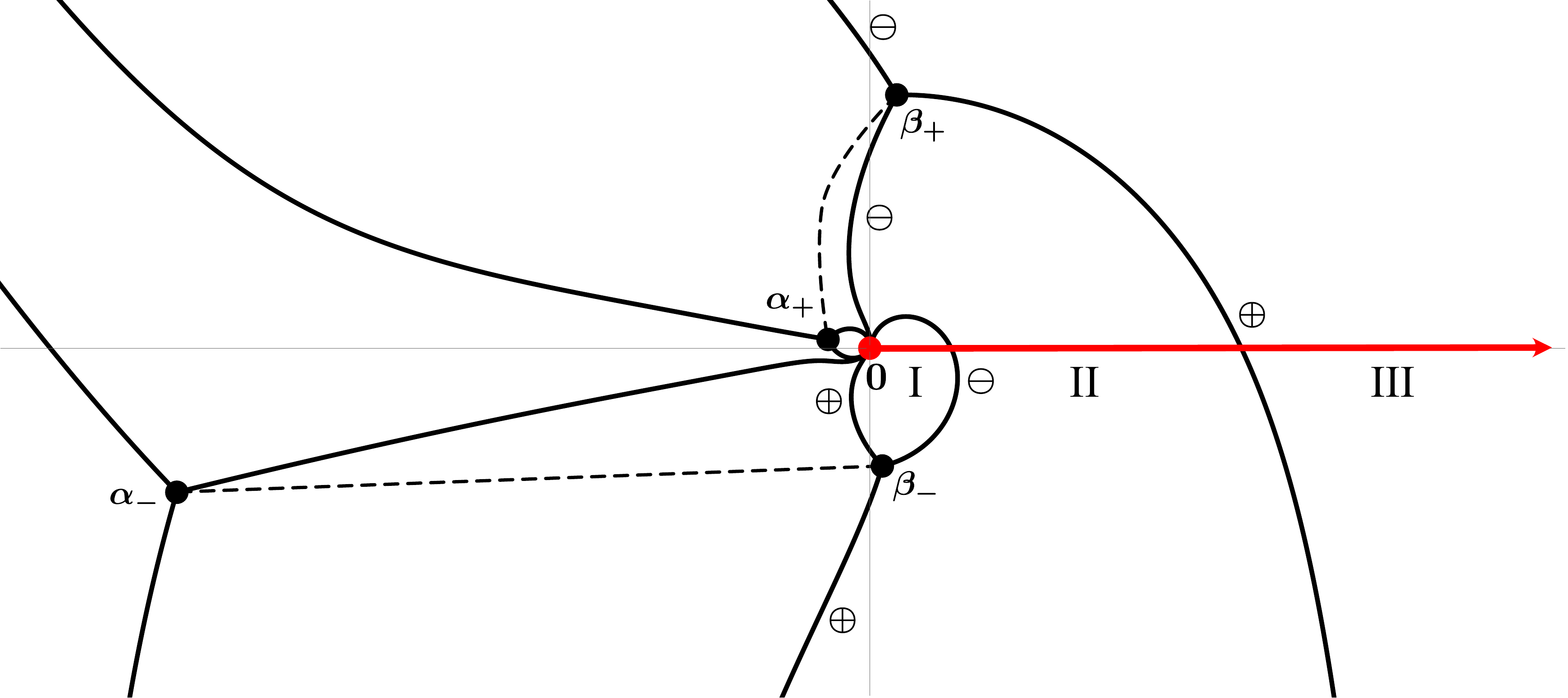}
\caption{Analytic continuation of the Borel resummed WKB solutions along the positive real axis from a neighborhood of $z=0$ to a neighborhood of $z=\infty$, using the connection formulae. The solid lines denote Stokes curves, while the dashed lines denote branch cuts. The symbols $\oplus$ and $\ominus$ indicate which of $\Psi_{\pm}$ is dominant on each Stokes curve.}
\label{fig:EQC}
\end{figure}

Now we can derive the EQC for this type of Stokes geometry. Along the positive real axis, we cross the two Stokes curves. We start with the Borel resummed WKB solution $\Psi_{\pm}^\mathrm{I}(z;\beta_-)$ in region I, as shown in Fig.~\ref{fig:EQC}. 
We have chosen the reference point $z=\beta_-$ because the first Stokes curve that we encounter emanates from this turning point. After crossing this Stokes curve, the dominant solution $\Psi_-^\mathrm{I}$ changes discontinuously due to the Stokes phenomenon.
As we have already seen, this change is precisely determined by the connection formula. Hence we have
\begin{align}
\begin{pmatrix}
\Psi_+^\mathrm{I}(z;\beta_-) \\
\Psi_-^\mathrm{I}(z;\beta_-)
\end{pmatrix}
=
\begin{pmatrix}
1 & 0 \\
-i & 1
\end{pmatrix}
\begin{pmatrix}
\Psi_+^\mathrm{II}(z;\beta_-) \\
\Psi_-^\mathrm{II}(z;\beta_-)
\end{pmatrix}.
\end{align}
We continue the solutions from region II to region III. To do so, we cross the Stokes curve emanating from another turning point $z=\beta_+$. To use the connection formula, we have to switch the reference points from $z=\beta_-$ to $z=\beta_+$.
This change causes an additional factor, called the Voros symbol. Explicitly, we have
\begin{align}
\begin{pmatrix}
\Psi_+^\mathrm{II}(z;\beta_-) \\
\Psi_-^\mathrm{II}(z;\beta_-)
\end{pmatrix}
=
\begin{pmatrix}
V_{\beta_- \beta_+}^{1/2} & 0 \\
0 & V_{\beta_- \beta_+}^{-1/2}
\end{pmatrix}
\begin{pmatrix}
\Psi_+^\mathrm{II}(z;\beta_+) \\
\Psi_-^\mathrm{II}(z;\beta_+)
\end{pmatrix},
\end{align}
where
\begin{equation}
\begin{aligned}
V_{ab}:=\exp \biggl[ \mathcal{S} \biggl(\oint_{\gamma_{ab}} S_\text{odd}(z) dz \biggr) \biggr].
\end{aligned}
\end{equation}
The integration path $\gamma_{ab}$ is a cycle encircling two turning points $a$ and $b$ counterclockwise.
Note that the Voros symbol should be understood as being defined by the Borel resummation of its formal expansion in $\hbar$.
Now we can continue the solutions from region II to region III. The result is
\begin{align}
\begin{pmatrix}
\Psi_+^\mathrm{II}(z;\beta_+) \\
\Psi_-^\mathrm{II}(z;\beta_+)
\end{pmatrix}
=
\begin{pmatrix}
1 & i \\
0 & 1
\end{pmatrix}
\begin{pmatrix}
\Psi_+^\mathrm{III}(z;\beta_+) \\
\Psi_-^\mathrm{III}(z;\beta_+)
\end{pmatrix}.
\end{align}
Combining all of these results, we finally obtain the analytic continuation from region I to region III:
\begin{align}
\begin{pmatrix}
\Psi_+^\mathrm{I}(z;\beta_-) \\
\Psi_-^\mathrm{I}(z;\beta_-)
\end{pmatrix}
=
\begin{pmatrix}
1 & 0 \\
-i & 1
\end{pmatrix}
\begin{pmatrix}
V_{\beta_- \beta_+}^{1/2} & 0 \\
0 & V_{\beta_- \beta_+}^{-1/2}
\end{pmatrix}
\begin{pmatrix}
1 & i \\
0 & 1
\end{pmatrix}
\begin{pmatrix}
\Psi_+^\mathrm{III}(z;\beta_+) \\
\Psi_-^\mathrm{III}(z;\beta_+)
\end{pmatrix}.%\\
%&=\begin{pmatrix}
%V_{\beta_- \beta_+}^{1/2} & iV_{\beta_- \beta_+}^{1/2} \\
%-iV_{\beta_- \beta_+}^{1/2} & V_{\beta_- \beta_+}^{1/2}+V_{\beta_- \beta_+}^{-1/2}
%\end{pmatrix}
%\begin{pmatrix}
%\Psi_+^\mathrm{III}(x;\beta_+) \\
%\Psi_-^\mathrm{III}(x;\beta_+)
%\end{pmatrix}
\end{align}
In particular, we obtain
\begin{equation}
\begin{aligned}
\Psi_-^\mathrm{I}(z;\beta_-)=-iV_{\beta_- \beta_+}^{1/2}\Psi_+^\mathrm{III}(z;\beta_+)
+(V_{\beta_- \beta_+}^{1/2}+V_{\beta_- \beta_+}^{-1/2}) \Psi_-^\mathrm{III}(z;\beta_+).
\end{aligned}
\end{equation}
The QNM boundary condition requires that the coefficient of $\Psi_-^\mathrm{III}(z;\beta_+)$ in this equation must vanish.
Therefore we arrive at the exact condition
\begin{equation}
\begin{aligned}
V_{\beta_- \beta_+}=-1,
\end{aligned}
\end{equation}
which leads to the EQC:
\begin{equation}
\begin{aligned}
\mathcal{S} \biggl( \oint_{\gamma_{\beta_- \beta_+}} S_\text{odd}(z) dz \biggr)=2\pi i\biggl( n+\frac{1}{2} \biggr),\qquad
n \in \mathbb{Z}.
\end{aligned}
\label{eq:EQC}
\end{equation}
The left-hand side is referred to as the (Borel resummed) quantum period.
In the next subsection, we will discuss how to treat this EQC.

\subsection{Quantum Periods and QNM Frequencies}
To solve the EQC \eqref{eq:EQC}, we have to evaluate the Borel resummation of the quantum period for the cycle $\gamma_{\beta_- \beta_+}$. For this purpose, we recall the formal power series \eqref{eq:S_odd/even}. This leads to the formal power series of the quantum period,
\begin{equation}
\begin{aligned}
\oint_{\gamma_{\beta_- \beta_+}} S_\text{odd}(z) dz=\sum_{k=0}^\infty \hbar^{2k-1} \oint_{\gamma_{\beta_- \beta_+}} S_{2k-1}(z) dz,
\end{aligned}
\end{equation}
by interchanging the summation and the integration.
Our strategy is to evaluate 
\begin{equation}
\begin{aligned}
\oint_{\gamma_{\beta_- \beta_+}} S_{2k-1}(z) dz, \qquad k=0,1,2,\dots,
\end{aligned}
\end{equation}
order by order and then perform the Borel resummation of the resulting formal power series using Pad\'e approximants.
The analytic evaluation of the quantum periods is not simple in general. However, in the extremal case, it can be done up to quite large $k$. We have obtained the analytic data up to $k=160$. The computation becomes more involved in the non-extremal case or for gravitational and electromagnetic perturbations.

We now present our analytic results for the quantum period.
We notice that the identification \eqref{eq:id-1} is not the only choice of the parameters because there is a rescaling freedom
\begin{equation}
\begin{aligned}
(\hbar, \Lambda, m_1, m_2, u) \to (\kappa \hbar, \kappa \Lambda, \kappa m_1, \kappa m_2, \kappa^2 u),
\end{aligned}
\label{eq:rescale}
\end{equation}
with an arbitrary non-zero parameter $\kappa$.
For later purposes, it is more convenient to choose $\kappa=\Lambda^{-1}$, which gives
\begin{equation}
\begin{aligned}
\hbar=\frac{i}{M\omega},\qquad 2u=-6+\frac{(\ell+1/2)^2}{(M\omega)^2},\qquad
\Lambda=1,\qquad m_1=m_2=2.
\end{aligned}
\end{equation}
We also introduce
%\begin{equation}
%\begin{aligned}
%E=\frac{(\ell+1/2)^2}{(M\omega)^2},
%\end{aligned}
%\end{equation}
%and
\begin{equation}
\begin{aligned}
\Pi_B^{(k)} :=i \oint_{\gamma_{\beta_- \beta_+}} S_{2k-1}(z) dz.
\end{aligned}
\end{equation}
The detailed computation of $\Pi_B^{(k)}$ is explained in Appendix~\ref{app:ACQP}. Here we show the final results.
The classical period is written, in terms of the complete elliptic integrals, as
\begin{equation}
\begin{aligned}
\Pi_B^{(0)}=\frac{2}{\sqrt{A(4A+1)}} \biggl[ 4A(8A+1)\eK\biggl( \frac{4A-1}{4A+1} \biggr)
-4A(4A+1)\eE\biggl( \frac{4A-1}{4A+1} \biggr)\\
-\ePi\biggl( 1-\frac{1}{4A} \bigg| \frac{4A-1}{4A+1} \biggr)\biggr],
\end{aligned}
\end{equation}
where $A=(\ell+1/2)/(M\omega)$, and the complete elliptic integrals are defined by
\begin{align}
\eK(\mathsf{m})&:=\int_0^{\pi/2} \frac{d\theta}{\sqrt{1-\mathsf{m}\sin^2 \theta}}=\int_0^1 \frac{dt}{\sqrt{(1-t^2)(1-\mathsf{m}\,t^2)}}, \label{eq:eK}\\
\eE(\mathsf{m})&:=\int_0^{\pi/2} \sqrt{1-\mathsf{m}\sin^2 \theta}d\theta=\int_0^1 \sqrt{\frac{1-\mathsf{m}\,t^2}{1-t^2}}dt, \label{eq:eE}\\
\ePi(\mathsf{n}|\mathsf{m})&:=\int_0^{\pi/2} \frac{d\theta}{(1-\mathsf{n} \sin^2 \theta)\sqrt{1-\mathsf{m}\sin^2 \theta}}. \label{eq:ePi}
\end{align}
The quantum corrections for $k=1,2$ are also given by
\begin{align}
\Pi_B^{(1)}&=\frac{1}{96A^{\frac{3}{2}}(4A-1)\sqrt{4A+1}}\biggl[ (16A-3)\eK\biggl( \frac{4A-1}{4A+1} \biggr)
-(64A^2-3)\eE\biggl( \frac{4A-1}{4A+1} \biggr) \biggr], \\
\Pi_B^{(2)}&=\frac{1}{11796480A^{\frac{9}{2}}(4A-1)^3(4A+1)^{\frac{5}{2}}}\biggl[ (1048576 A^6-196608 A^5-184320 A^4\notag \\
&\quad+9600 A^3+9168 A^2-504 A-165)\eK\biggl( \frac{4A-1}{4A+1} \biggr)\notag \\
&\quad-(4194304 A^7-933888 A^5+46272 A^3-1164 A)\eE\biggl( \frac{4A-1}{4A+1} \biggr) \biggr].
\end{align}
In general, we observe that, for $k \geq 1$, the quantum corrections take the following form: 
\begin{equation}
\begin{aligned}
\Pi_B^{(k)}=\frac{1}{A^{3(k-\frac{1}{2})}(4A-1)^{2k-1}(4A+1)^{2k-\frac{3}{2}}} \biggl[ P_{\mathrm{K}}^{(k)} \eK\biggl( \frac{4A-1}{4A+1} \biggr)+ P_{\mathrm{E}}^{(k)} \eE\biggl( \frac{4A-1}{4A+1} \biggr) \biggr],
\end{aligned}
\end{equation}
where $P_{\mathrm{K}}^{(k)}$ and $P_{\mathrm{E}}^{(k)}$ are polynomials of degrees $5k-4$ and $5k-3$ in $A$, respectively.
As mentioned before, we have determined these polynomials up to $k=160$.

\begin{figure}[tb]
  \begin{minipage}[b]{0.32\linewidth}
    \centering
    \includegraphics[width=0.95\linewidth]{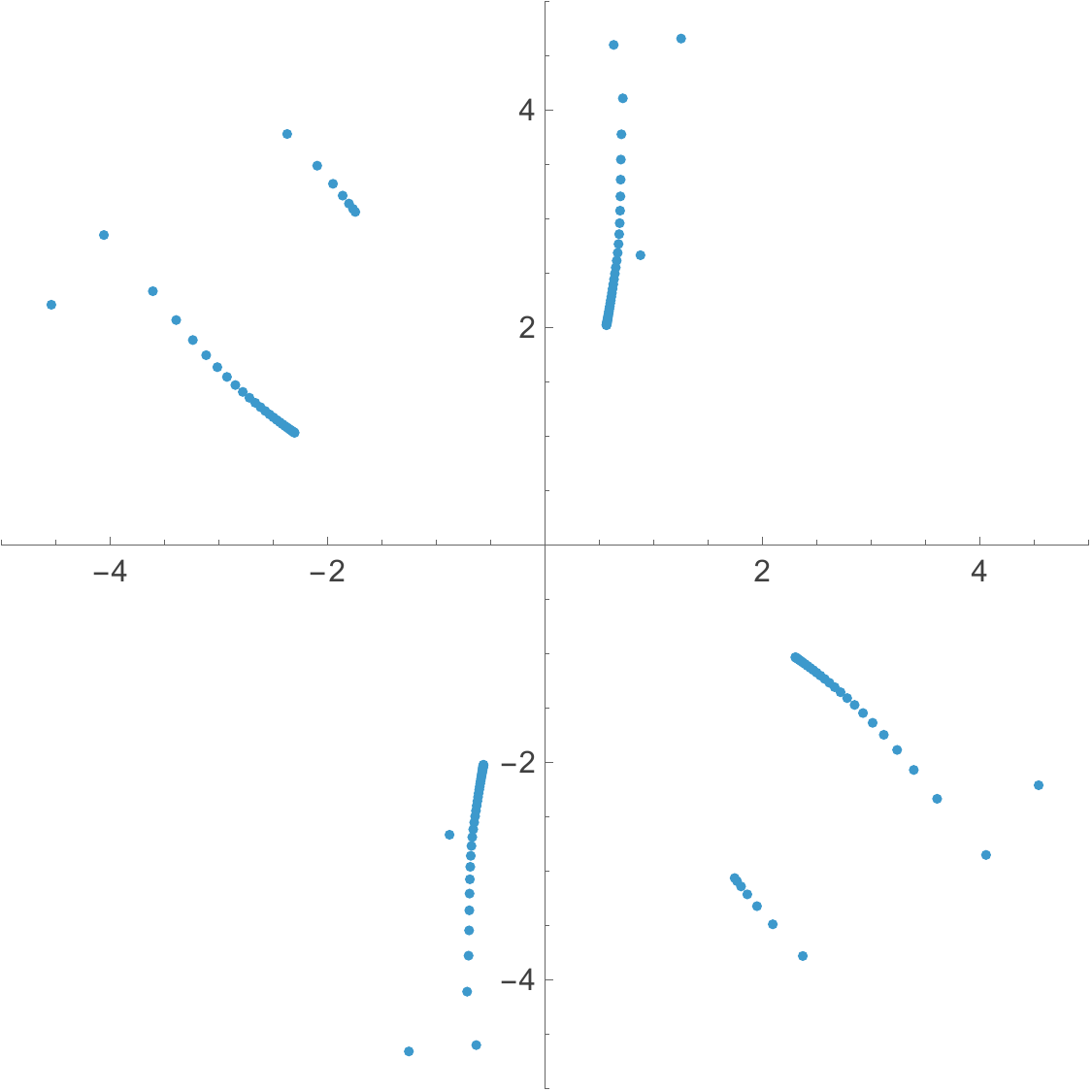}\\
    {\footnotesize (a) $\ell=0$ and $M\omega=0.1-0.1i$.}
  \end{minipage}
    \begin{minipage}[b]{0.32\linewidth}
    \centering
    \includegraphics[width=0.97\linewidth]{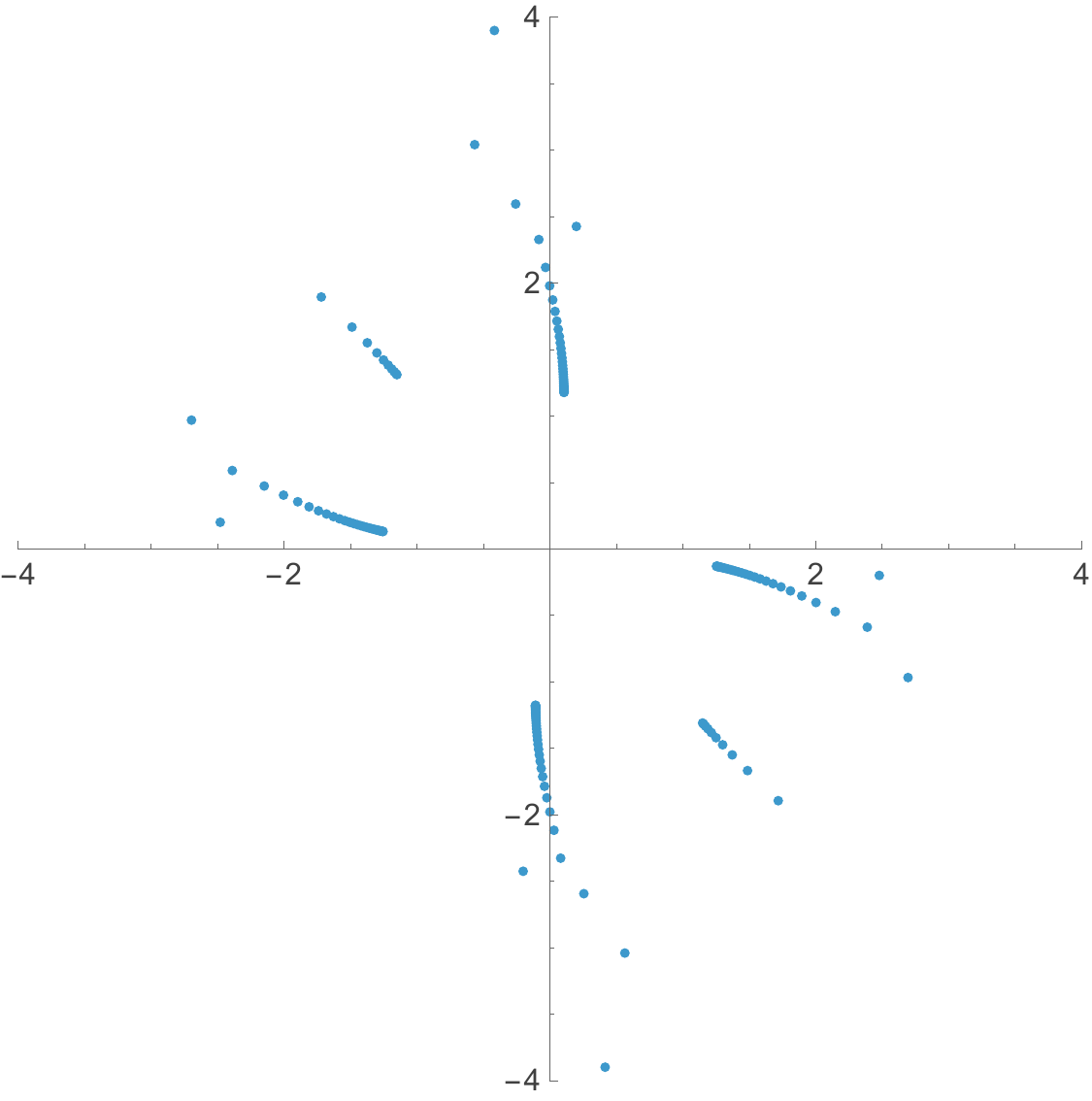}\\
    {\footnotesize (b) $\ell=0$ and $M\omega=0.1-0.5i$.}
  \end{minipage}
      \begin{minipage}[b]{0.32\linewidth}
    \centering
    \includegraphics[width=0.95\linewidth]{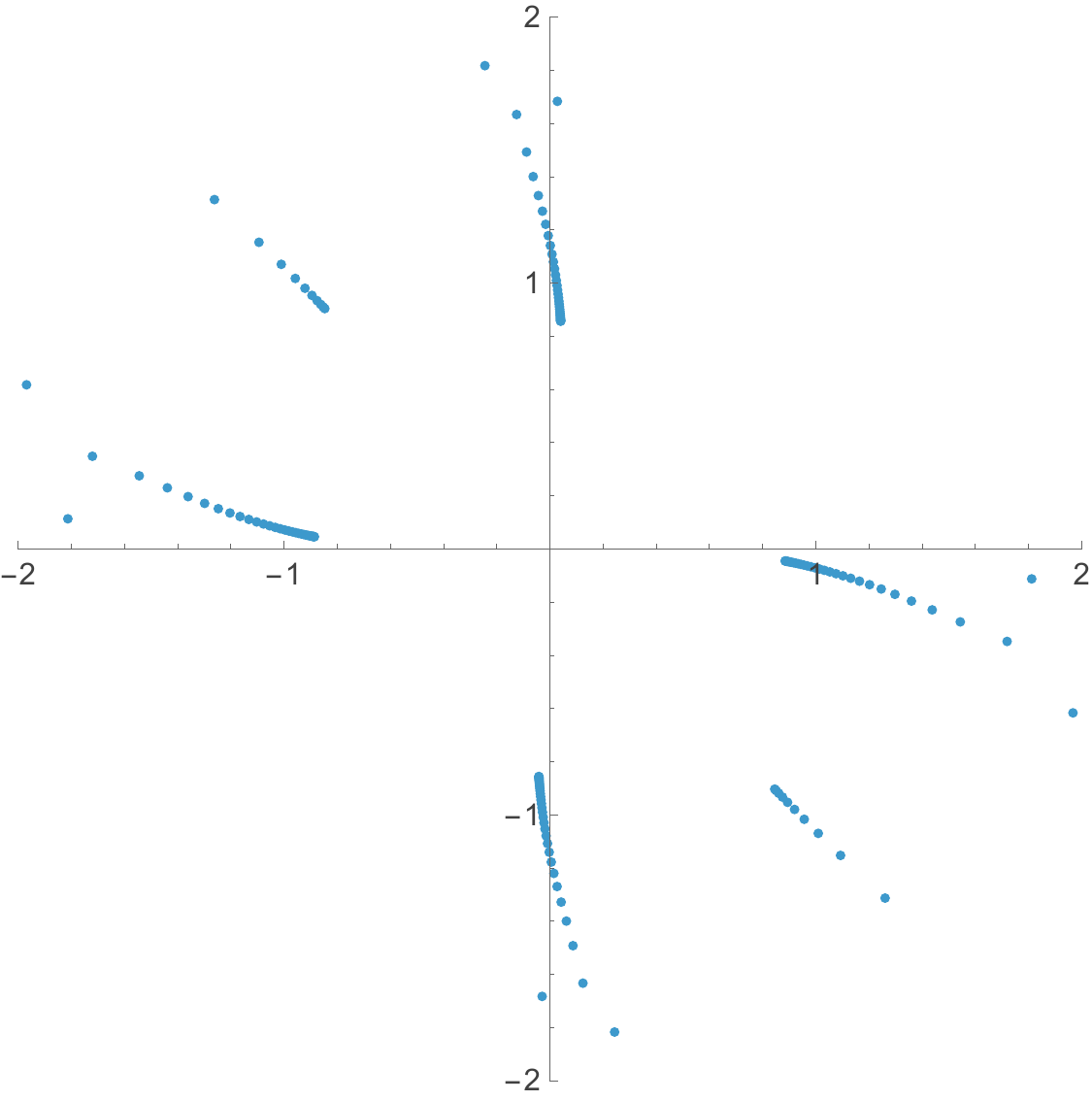}\\
    {\footnotesize (c) $\ell=0$ and $M\omega=0.1-i$.}
  \end{minipage}
  \caption{Pole distribution of the Pad\'e approximant $\widehat{\Pi}_B^{[160/160]}(\hbar \zeta)$.}
  \label{fig:Borel-sing}
\end{figure}

The EQC is now given by
\begin{equation}
\begin{aligned}
\mathcal{S} (\Pi_B)=2\pi \hbar \biggl( n+\frac{1}{2} \biggr),\qquad
\Pi_B:=\sum_{k=0}^\infty \hbar^{2k} \Pi_B^{(k)}.
\end{aligned}
\end{equation}
We proceed to the Borel resummation of the quantum period, which is explicitly given by
\begin{equation}
\begin{aligned}
\mathcal{S} (\Pi_B)
=\int_{0}^{\infty} e^{-\zeta} \widehat{\Pi}_B (\hbar \zeta) d\zeta,
\end{aligned}
\end{equation}
with the Borel transform,
\begin{equation}
\begin{aligned}
\widehat{\Pi}_B (\zeta)=\sum_{k=0}^\infty \frac{\Pi_B^{(k)}}{(2k)!}\zeta^{2k}. 
\end{aligned}
\end{equation}
In practical computations, one only has access to a finite set of perturbative data $\Pi_B^{(k)}$ up to some finite order $k$.
In such cases, we approximate the Borel transform by its Pad\'e approximants.
Let $F^{[M/N]}(x)$ denote the Pad\'e approximant of $F(x)$ whose numerator and denominator are polynomials of degrees $M$ and $N$, respectively. Then the Borel--Pad\'e resummation is given by
\begin{equation}
\begin{aligned}
\mathcal{S}^{[M/N]} (\Pi_B)
=\int_{0}^{\infty} e^{-\zeta} \widehat{\Pi}_B^{[M/N]} (\hbar \zeta) d\zeta.
\end{aligned}
\end{equation}
Note that, in order to define the Borel resummation along the positive real axis, the Borel transform should have no singularities on this integration contour.
The singularity structure of the Borel transform $\widehat{\Pi}_B(\hbar \zeta)$ can be probed by examining the pole structure of its Pad\'e approximant $\widehat{\Pi}_B^{[M/N]}(\hbar \zeta)$.
In Fig.~\ref{fig:Borel-sing}, we show the pole structure of $\widehat{\Pi}_B^{[160/160]}(\hbar \zeta)$ for $\ell=0$ and $M\omega=0.1-0.1i$, $0.1-0.5i$, $0.1-i$.
These plots indicate that there are no singularities on the positive real axis.
They also suggest that, as $|\operatorname{Im}(M\omega)|$ increases, the nearest singularities approach the real and imaginary axes.

We also introduce a quantity
\begin{equation}
\begin{aligned}
\Delta_n^{[M/N]}:=\frac{\mathcal{S}^{[M/N]} (\Pi_B)}{2\pi \hbar} -\biggl(n+\frac{1}{2}\biggr).
\end{aligned}
\end{equation}
For given $\ell$ and $n$, we can solve the Borel--Pad\'e resummed quantization condition $\Delta_n^{[M/N]}=0$ by Newton's method, which provides us with an approximate value of the QNM frequency $M\omega_{n}^{\ell}$.
To test our EQC, we use a less computationally demanding approach.
First, we take the values of the QNM frequencies, obtained by an independent method, with sufficiently high precision. 
Next, we substitute these values into $\Delta_n^{[M/N]}$. By examining how far the result deviates from zero, we can assess the accuracy of our Borel--Pad\'e resummed quantization condition. 

For example, for $(\ell,n)=(0,0)$, we have the following highly accurate QNM frequency:\footnote{We obtained this value by using the method in~\cite{Onozawa:1995vu, Richartz:2015saa}. We also thank Masashi Kimura for sharing the QNM frequencies with extremely high precision, which can be compared with ours independently.}
\begin{equation}
\begin{aligned}
M\omega_0^{\ell=0}&=0.13345889356706911632231170942642298452384071180138\\ 
&\quad-0.09584384212811504931055969700757251219004718021629i.
\end{aligned}
\end{equation}
Substituting it into $\Delta_0^{[160/160]}$, we obtain
\begin{equation}
\begin{aligned}
\bigl|\Delta_0^{[160/160]}(\omega=\omega_0^{\ell=0})\bigr| \approx 3.38 \times 10^{-10}.
\end{aligned}
\end{equation}
Therefore the Borel--Pad\'e quantization condition has roughly 10-digit precision for $\ell=0$ and $n=0$.

\begin{figure}[tb]
  \begin{minipage}[b]{0.45\linewidth}
    \centering
    \includegraphics[width=0.95\linewidth]{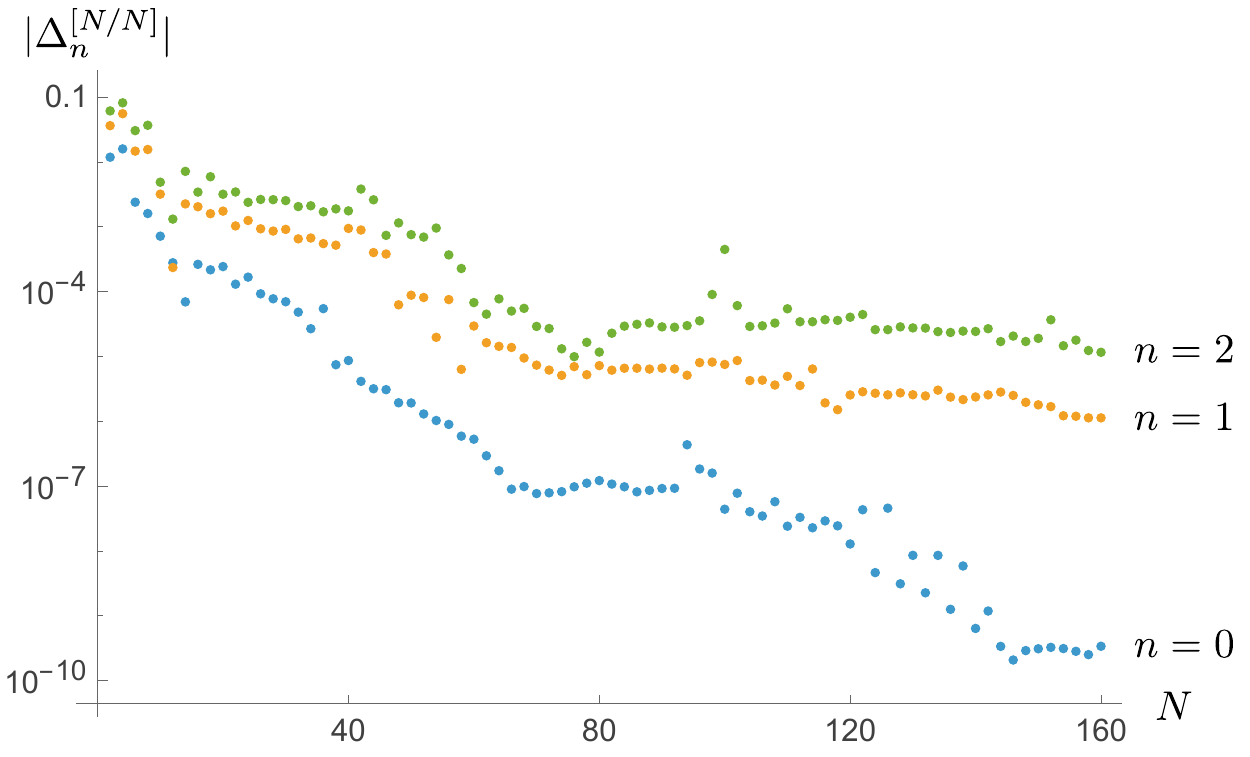}\\
    {\footnotesize (a) $\ell=0$.}
  \end{minipage}
    \begin{minipage}[b]{0.45\linewidth}
    \centering
    \includegraphics[width=0.95\linewidth]{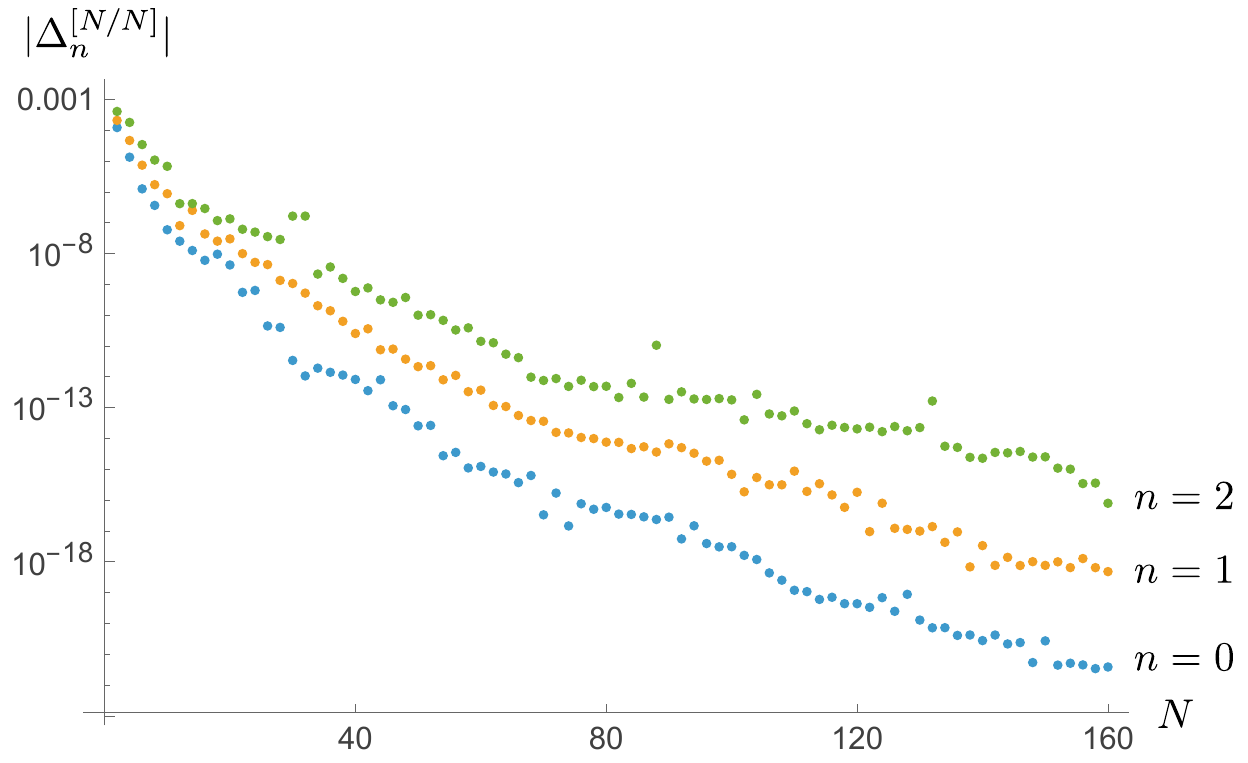}\\
    {\footnotesize (b) $\ell=1$.}
  \end{minipage} \\
    \begin{minipage}[b]{0.45\linewidth}
    \centering
    \includegraphics[width=0.95\linewidth]{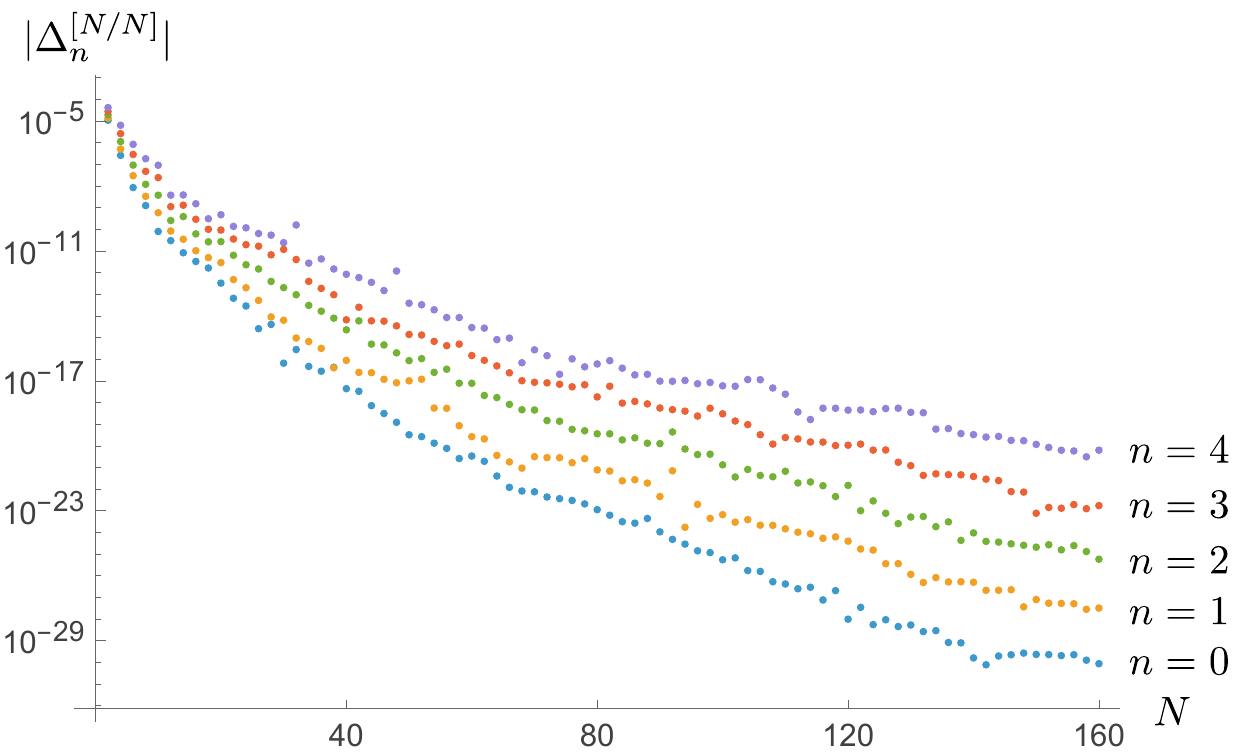}\\
    {\footnotesize (c) $\ell=2$.}
  \end{minipage}
  \begin{minipage}[b]{0.45\linewidth}
    \centering
    \includegraphics[width=0.95\linewidth]{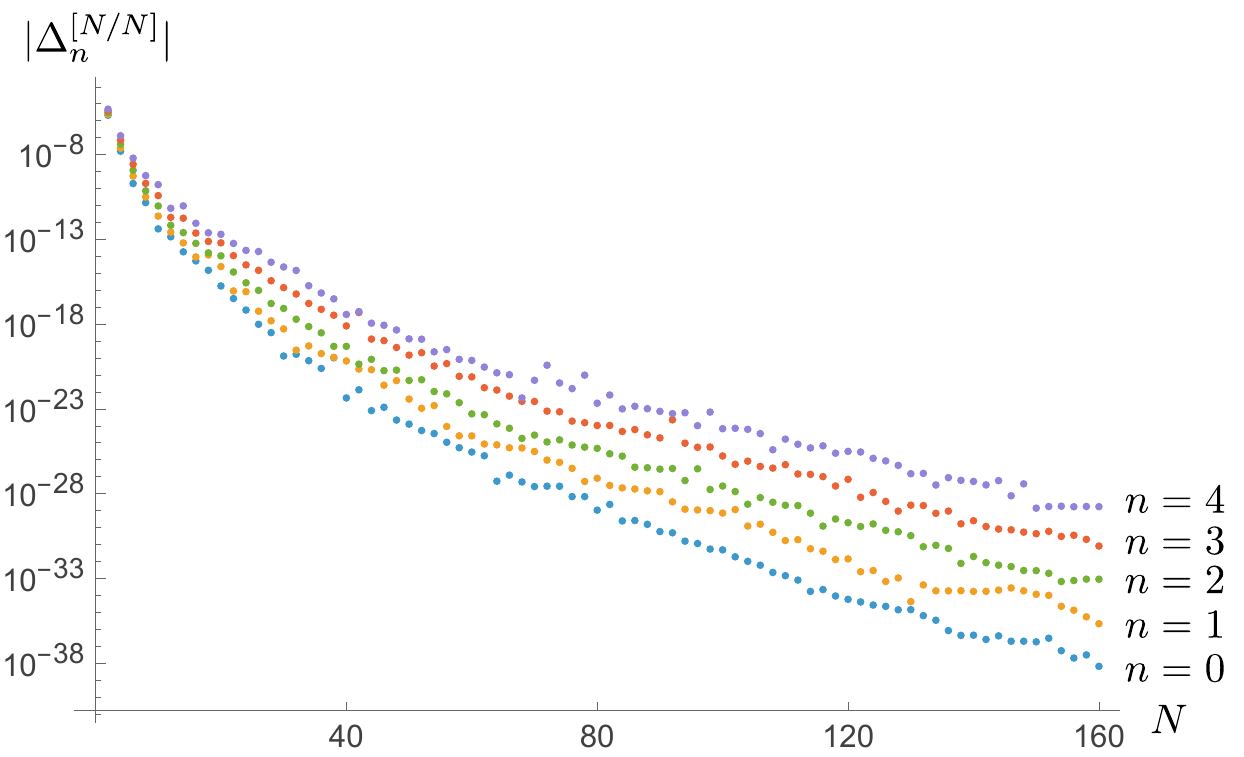}\\
    {\footnotesize (d) $\ell=3$.}
  \end{minipage}
  \caption{These graphs test the Borel--Pad\'e resummed quantization condition for the extremal RN black hole. We substitute highly accurate QNM frequencies into $\Delta_n^{[N/N]}$. The deviation of $|\Delta_n^{[N/N]}|$ from zero provides a measure of the accuracy of the quantization condition.}
  \label{fig:Delta}
\end{figure}

Figure~\ref{fig:Delta} shows the magnitude of the deviation of the diagonal Borel--Pad\'e resummed quantity $\Delta_n^{[N/N]}$ from zero for $\ell=0,1,2,3$. One can see that $|\Delta_n^{[N/N]}|$ approaches zero as $N$ increases. As in the conventional WKB analysis, the accuracy of the approximation improves as $\ell$ increases.

\section{Extremal Kerr Black Hole}
In this section, we perform an analogous analysis for the extremal Kerr black hole. The computation is straightforward but technically more involved. We begin with the Kerr metric in Boyer--Lindquist coordinates,
\begin{equation}
\begin{aligned}
ds^2=-\frac{\Delta}{\Sigma}(dt-a \sin^2 \theta d\phi)^2+\frac{\Sigma}{\Delta} dr^2+\Sigma d\theta^2+\frac{\sin^2 \theta}{\Sigma}\bigl( (r^2+a^2)d\phi-adt \big)^2,
\end{aligned}
\end{equation}
where
\begin{equation}
\begin{aligned}
\Sigma=r^2+a^2 \cos^2 \theta,\qquad \Delta=r^2-2Mr+a^2.
\end{aligned}
\end{equation}
The Klein--Gordon equation \eqref{eq:KG} on this background is known as the Teukolsky equation with $s=0$.
The Teukolsky equation can be separated by adopting the ansatz $\Phi = R(r)S(\theta)e^{-i\omega t-im\phi}$.
The resulting equations are given by
\begin{align}
&\biggl[ \frac{1}{\sin \theta} \frac{d}{d\theta} \biggl(\sin \theta \frac{d}{d\theta} \biggr)-(a\omega)^2 \sin^2 \theta-\frac{m^2}{\sin^2 \theta}+\lambda \biggr]S(\theta)=0, \label{eq:angular-Teukolsky} \\
&\biggl[ \frac{d}{dr}\biggl( \Delta \frac{d}{dr}\biggr)+\frac{K^2}{\Delta}+2ma\omega -\lambda \biggr]R(r)=0,\label{eq:radial-Teukolsky}
\end{align}
where
\begin{equation}
\begin{aligned}
K=(r^2+a^2)\omega-ma.
\end{aligned}
\end{equation}
For $0\leq a <M$, $\Delta=0$ has two distinct roots,
\begin{align}
r_{\pm}=M \pm \sqrt{M^2-a^2},
\end{align}
where $r=r_+$ corresponds to the event horizon.
The tortoise coordinate is defined by
\begin{align}
\frac{dr_*}{dr}=\frac{r^2+a^2}{\Delta},
\end{align}
%The limit $r_* \to -\infty$ corresponds to the event horizon, while $r_* \to +\infty$ corresponds to spatial infinity.
and the QNM boundary conditions are given by
\begin{align}
R(r) \to \begin{cases} e^{-i(\omega-m\Omega_H) r_*} \quad &(r_* \to -\infty) \\
e^{i\omega r_*} \quad &(r_* \to +\infty) \end{cases},\qquad
\Omega_H=\frac{a}{r_+^2+a^2}.
\label{eq:QNM-Kerr-1}
\end{align}
The angular part \eqref{eq:angular-Teukolsky} is nothing but the spheroidal equation. Its eigenvalues are obtained by requiring regularity of $S(\theta)$ at $\theta=0,\pi$. Since the eigenvalue has two indices $(\ell, m)$, we denote it by $\lambda_{\ell m}$.\footnote{This spheroidal eigenvalue is directly related to the corresponding Mathematica function. We easily find $\lambda_{\ell m}=\mathtt{SpheroidalEigenvalue[\ell, m, I\, a\, \omega]}$.
}
In the limit $a \to 0$, we have $\lambda_{\ell m}=\ell(\ell+1)+O(a)$.

\begin{figure}[tb]
  \begin{minipage}[b]{0.45\linewidth}
    \centering
    \includegraphics[width=0.95\linewidth]{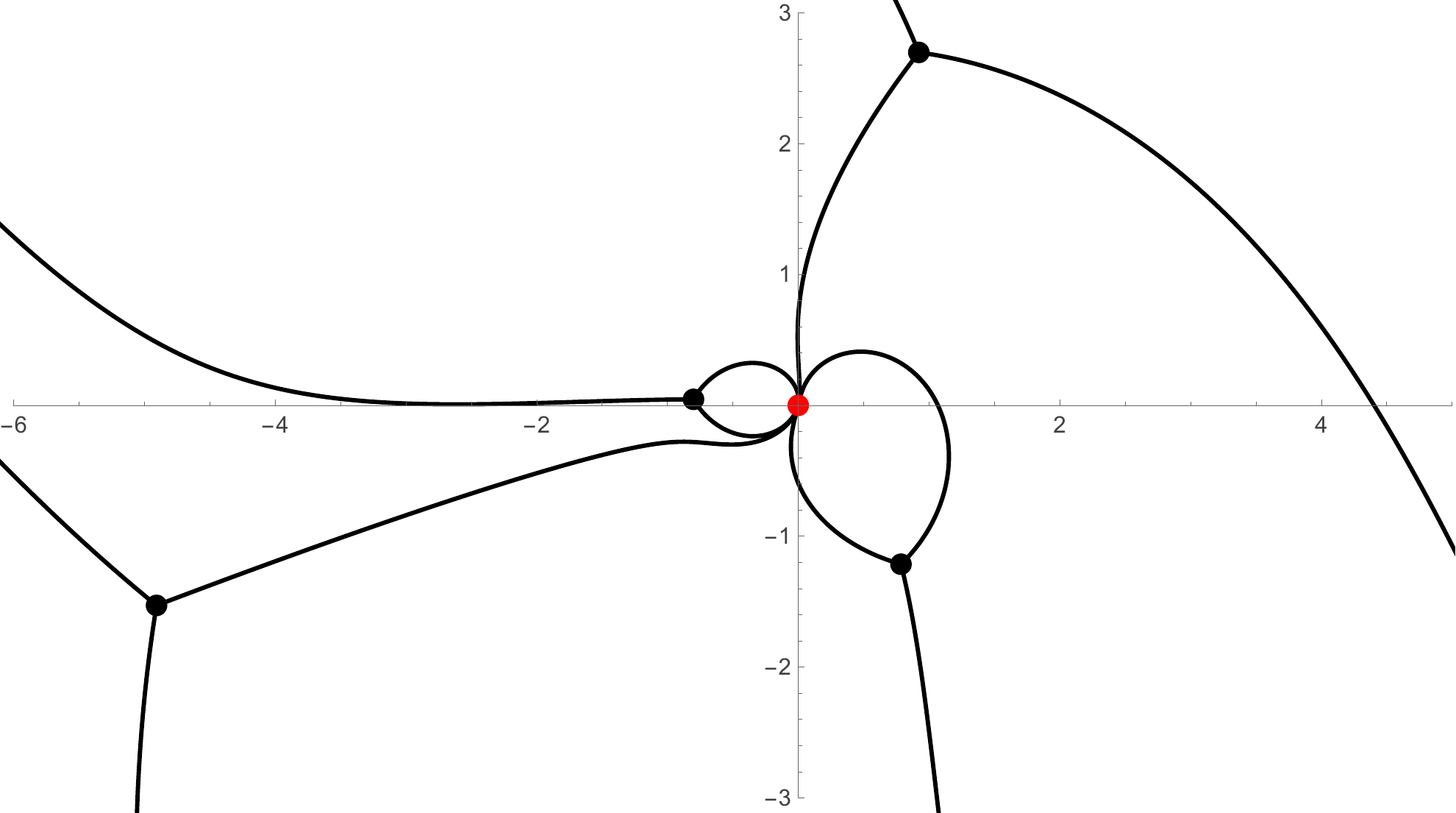}\\
    {\footnotesize (a) $(\ell,m)=(1,-1)$ and $M\omega=0.4-0.2i$.}
  \end{minipage}
    \begin{minipage}[b]{0.45\linewidth}
    \centering
    \includegraphics[width=0.95\linewidth]{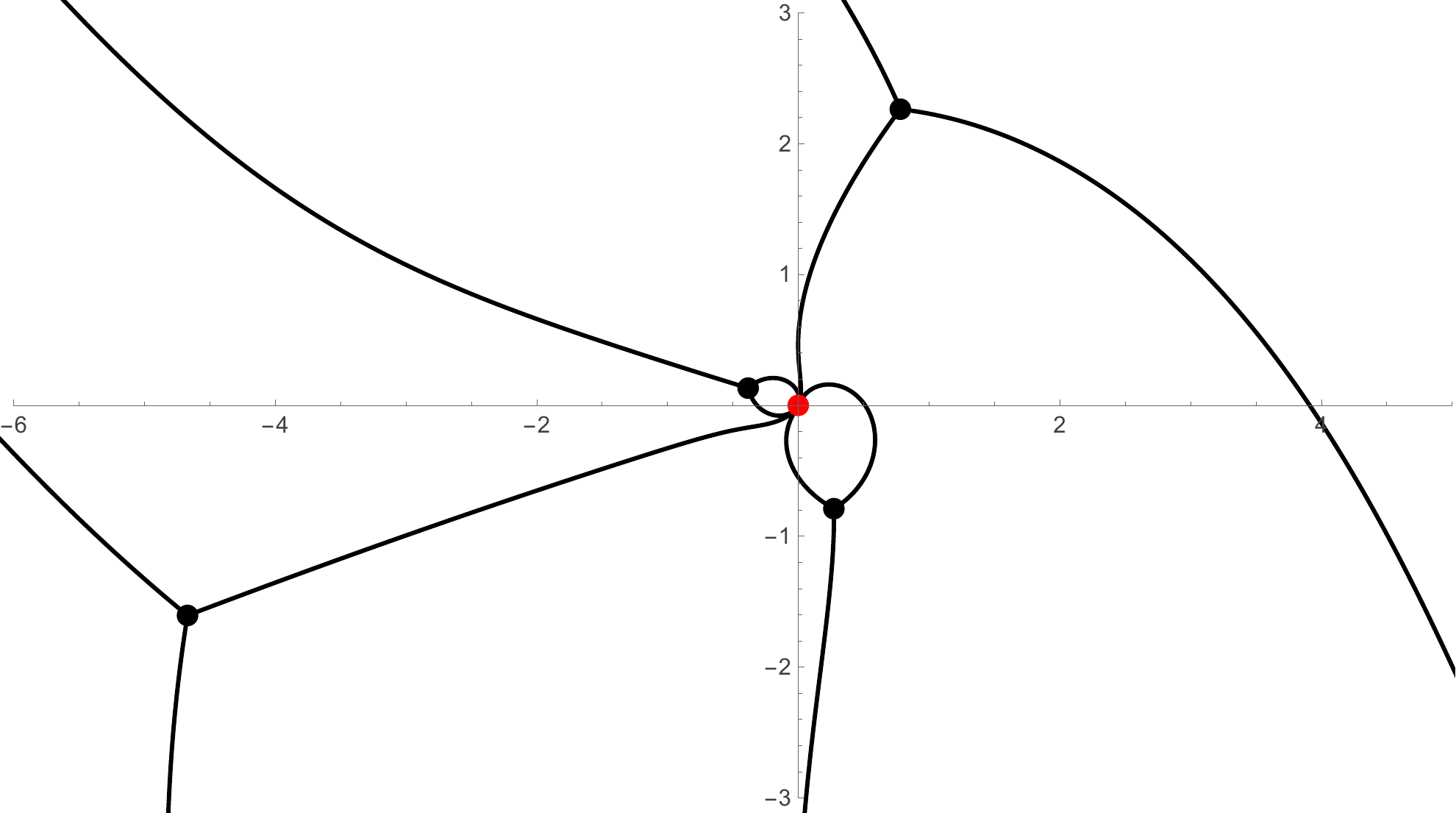}\\
    {\footnotesize (b) $(\ell,m)=(1,0)$ and $M\omega=0.4-0.2i$.}
  \end{minipage}\\
  \begin{minipage}[b]{0.45\linewidth}
    \centering
    \includegraphics[width=0.95\linewidth]{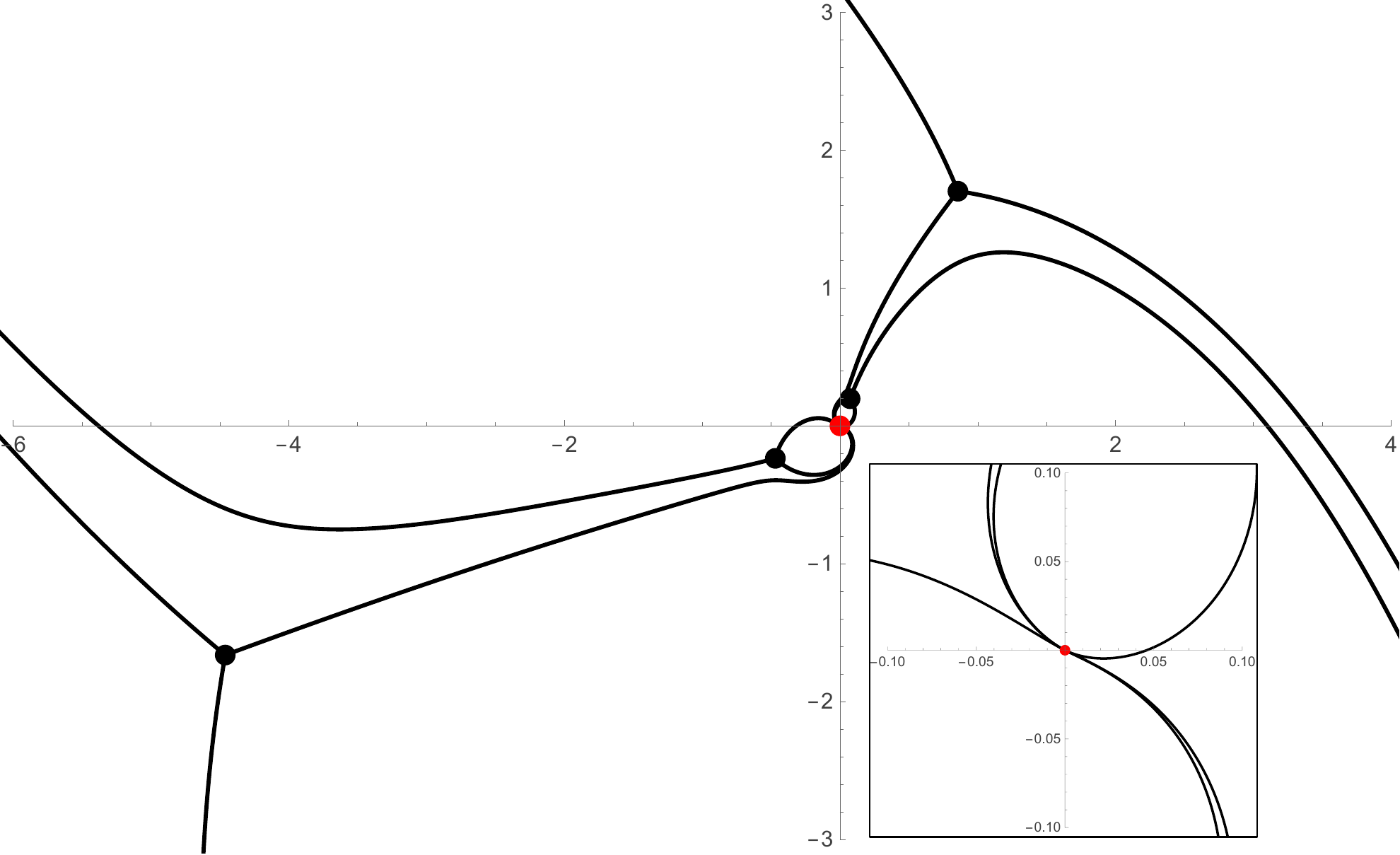}\\
    {\footnotesize (c) $(\ell,m)=(1,1)$ and $M\omega=0.4-0.2i$.}
  \end{minipage}
    \begin{minipage}[b]{0.45\linewidth}
    \centering
    \includegraphics[width=0.95\linewidth]{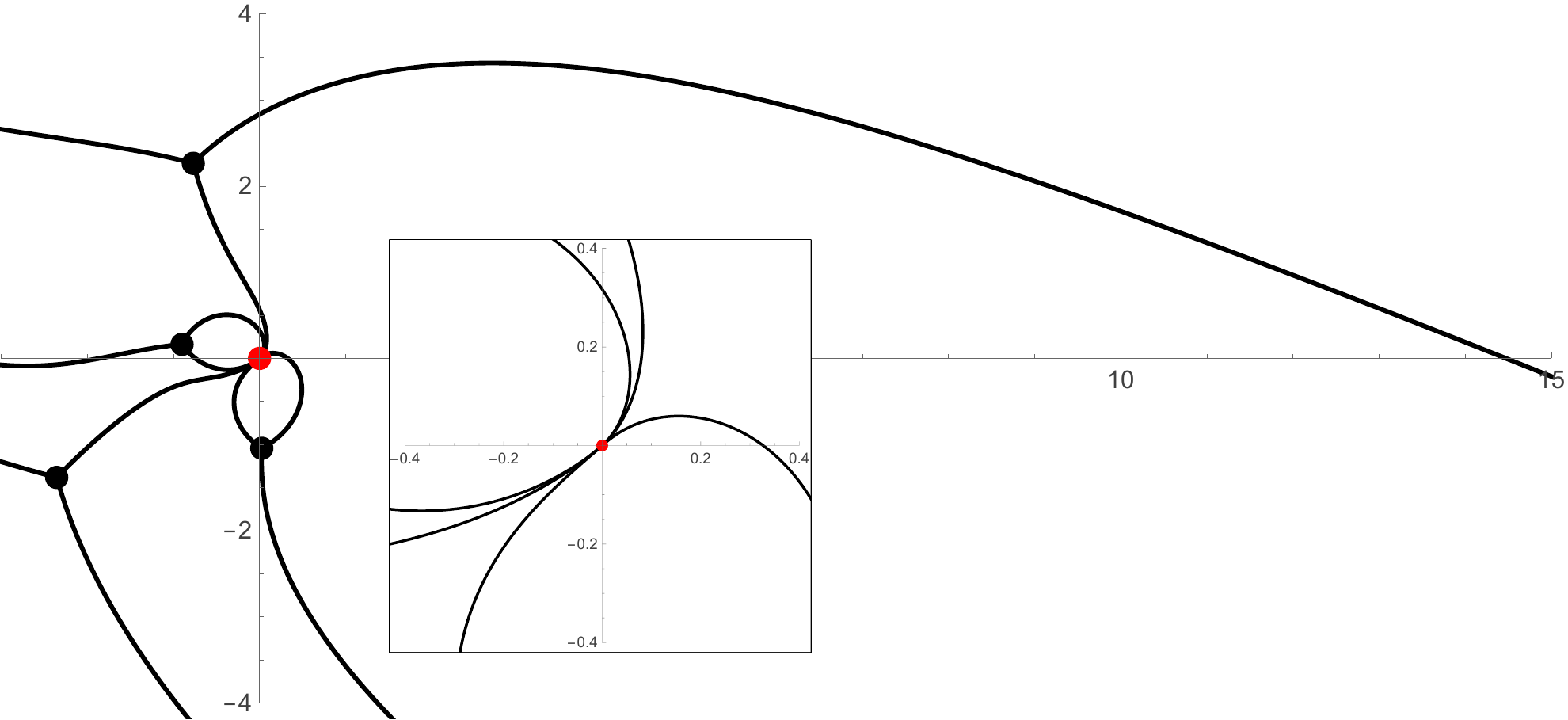}\\
    {\footnotesize (d) $(\ell,m)=(1,-1)$ and $M\omega=0.4-i$.}
  \end{minipage}\\
  \begin{minipage}[b]{0.45\linewidth}
    \centering
    \includegraphics[width=0.95\linewidth]{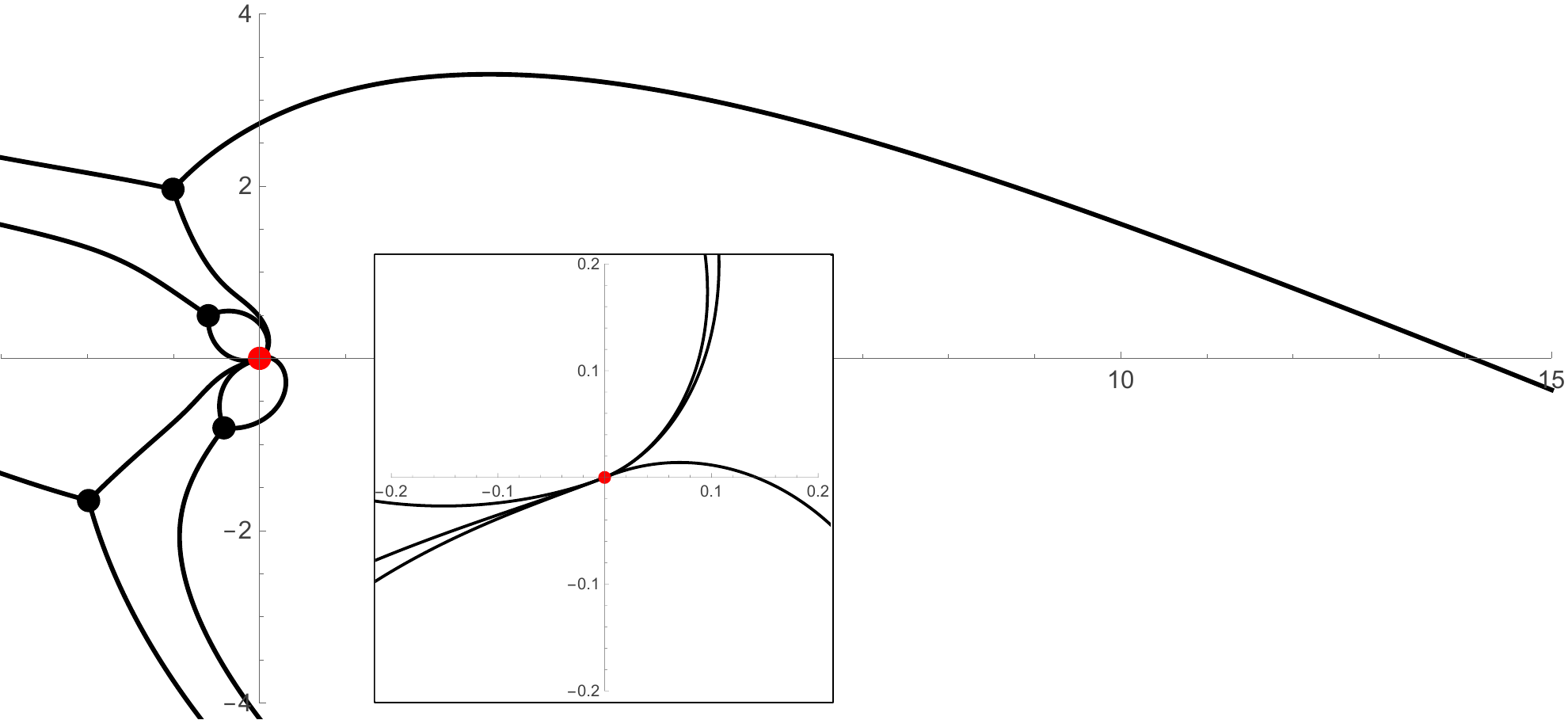}\\
    {\footnotesize (e) $(\ell,m)=(1,0)$ and $M\omega=0.4-i$.}
  \end{minipage}
    \begin{minipage}[b]{0.45\linewidth}
    \centering
    \includegraphics[width=0.95\linewidth]{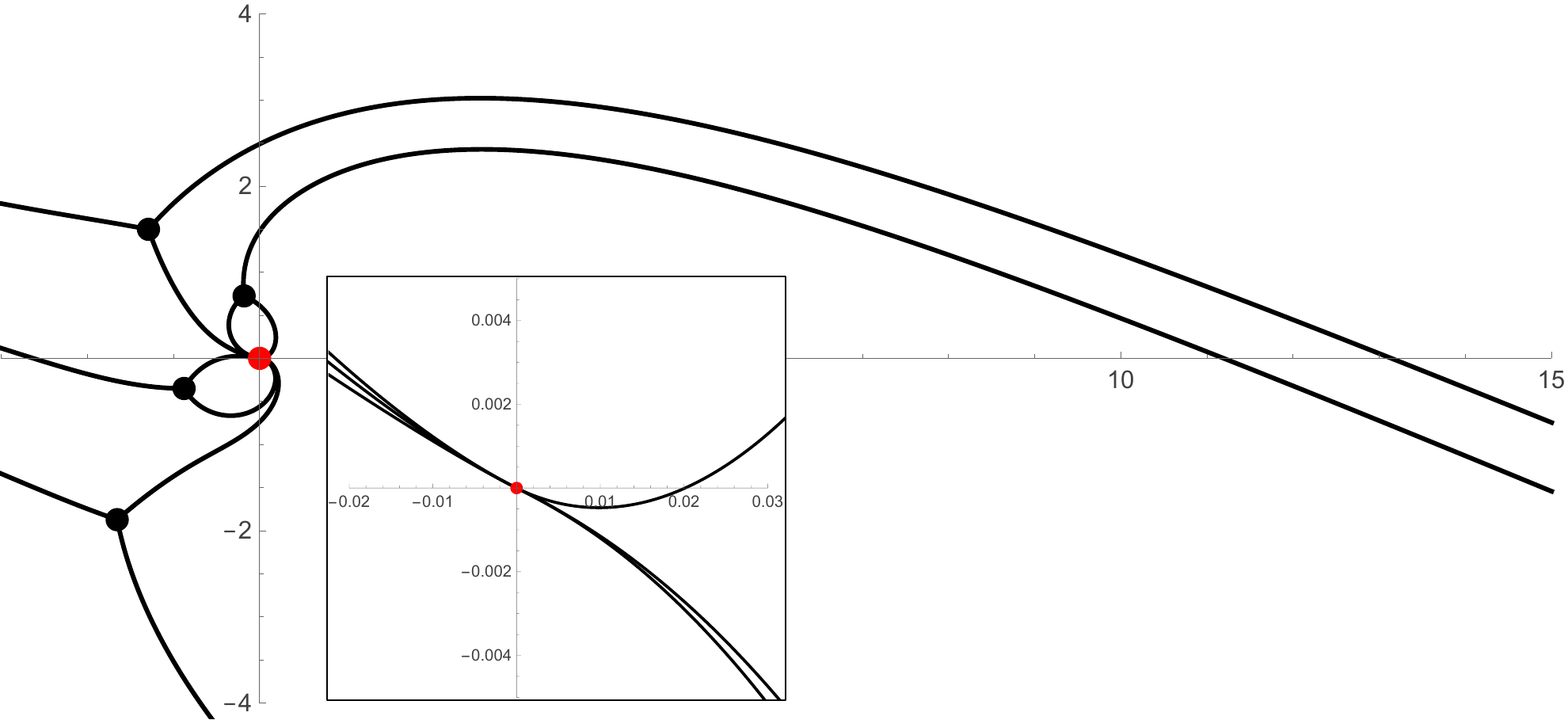}\\
    {\footnotesize (f) $(\ell,m)=(1,1)$ and $M\omega=0.4-i$.}
  \end{minipage}\\
  \caption{Stokes graphs of the extremal Kerr black hole for selected values of $(\ell,m)$ and $M\omega$. The Stokes geometry differs from the other cases only for $m=\ell>0$.}
  \label{fig:Stokes-Kerr}
\end{figure}

Now we take the extremal limit $a \to M$. In this limit, $\Delta=(r-M)^2$ has a double zero at $r=M$.
The tortoise coordinate is simply given by
\begin{align}
r_*=r-\frac{2M^2}{r-M}+2M \log(r-M)+\text{const}.
\end{align}
After changing the variables,
\begin{align}
r=M(1+z),\qquad R(r)=\frac{1}{z} \psi(z),
\end{align}
we obtain the Schr\"odinger-type equation \eqref{eq:Sch-2} with $\hbar=1$ and
\begin{align}
Q_0(z)&=-\frac{(2M\omega-m)^2}{z^4}-\frac{4M\omega(2M\omega-m)}{z^3}-\frac{8(M\omega)^2-\lambda_{\ell m}-1/4}{z^2}\notag \\
&\hspace{7truecm}-\frac{4(M\omega)^2}{z}-(M\omega)^2,\\
Q_2(z)&=-\frac{1}{4z^2}
\end{align}
The QNM condition is mapped to
\begin{align}
\psi(z) \to \begin{cases} e^{i(2M\omega-m)/z} z^{1-i(2M\omega-m)}  \quad &(z \to 0^+) \\
e^{iM\omega z}z^{2i M\omega} \quad &(z \to +\infty) \end{cases}.
\label{eq:QNM-Kerr-2}
\end{align}
We apply exact WKB analysis to this system. Examining Stokes graphs for representative parameter values, we observe that the Stokes geometry is essentially the same as in the RN case, except for $m=\ell>0$, as shown in Fig.~\ref{fig:Stokes-Kerr}.
The case $m=\ell>0$ is quite special because in this case the continuous extrapolation to the extremal limit $a \to M$ implies that the imaginary part of the QNM frequency goes to zero. It is unclear whether such a frequency satisfies the QNM condition~\cite{Richartz:2015saa}. We do not go into this subtle point in this paper.

Since the Stokes geometry is the same as in the RN case, the same EQC \eqref{eq:EQC} applies, and the remaining task is to evaluate the quantum period in the Kerr case.
To evaluate the quantum period, it is more convenient to rescale the variable as
\begin{align}
z \to \sqrt{\frac{2M\omega-m}{M\omega}}\, z.
\end{align}
After this rescaling, the potential functions in the Schr\"odinger equation take the form \eqref{eq:Q0Q2} with
\begin{equation}
\begin{aligned}
\hbar&=\frac{i}{\sqrt{M\omega(2M\omega-m)}},&\qquad 2u&=\frac{\lambda_{\ell m}+\frac{1}{4}-8(M\omega)^2}{M\omega(2M\omega-m)}, \\
\Lambda&=1,&\qquad \mu&:=m_1=m_2=\frac{2M\omega}{\sqrt{M\omega(2M\omega-m)}}.
\end{aligned}
\end{equation}
In particular, for $m=0$, the results become much simpler,
\begin{equation}
\begin{aligned}
\hbar=\frac{i}{\sqrt{2}M\omega},\qquad 2u=\frac{\lambda_{\ell 0}+\frac{1}{4}}{2(M\omega)^2}-4, \qquad
\Lambda=1,\qquad \mu=\sqrt{2}.
\end{aligned}
\end{equation}
The four turning points are also mapped to
\begin{align}
\alpha_\pm &=\frac{-A-\mu\pm \sqrt{(A+\mu-2)(A+\mu+2)}}{2} ,\\
\beta_\pm &=\frac{A-\mu \pm \sqrt{(A-\mu-2)(A-\mu+2)}}{2} ,
\end{align}
where we have defined
\begin{align}
A:=\sqrt{2u+2\Lambda^2+\mu^2}=\sqrt{\frac{\lambda_{\ell m}+1/4-2mM\omega}{M\omega(2M\omega-m)}}.
\end{align}
The quantum periods in this system are evaluated exactly, as explained in Appendix~\ref{app:ACQP}.
The classical period, for example, is given by
\begin{align}
\Pi_B^{(0)}&=8\sqrt{(-\alpha+1)(\beta+1)}\biggl[
-\eE(\mathsf{m})+\biggl(1+\frac{\mathsf{m}}{\mathsf{n}} \biggr) \eK(\mathsf{m})+\biggl(\mathsf{n}-\frac{\mathsf{m}}{\mathsf{n}} \biggr) \ePi(\mathsf{n}|\mathsf{m}) \biggr],
\label{eq:PiB0}
\end{align}
where
\begin{align}
\alpha:=\frac{\alpha_++\alpha_-}{2}=\frac{-A-\mu}{2},\qquad \beta:=\frac{\beta_++\beta_-}{2}=\frac{A-\mu}{2},
\end{align}
and
\begin{align}
\mathsf{m}=\frac{(-\alpha-1)(\beta-1)}{(-\alpha+1)(\beta+1)},\qquad \mathsf{n}=\frac{\beta-1}{\beta+1}.
\end{align}
The quantum periods take the following form:
\begin{align}
\Pi_B^{(k)}= B_K^{(k)} \eK( \mathsf{m} )+ B_E^{(k)} \eE( \mathsf{m} ),\qquad k=1,2,\dots,
\end{align}
where $B_K^{(k)}$ and $B_E^{(k)}$ are algebraic functions of $A$.
We have computed these quantum corrections up to $k=40$ for $\mu=\sqrt{2}$ ($m=0$) and $k=24$ for general $\mu$.

\begin{figure}[tb]
  \begin{minipage}[b]{0.45\linewidth}
    \centering
    \includegraphics[width=0.95\linewidth]{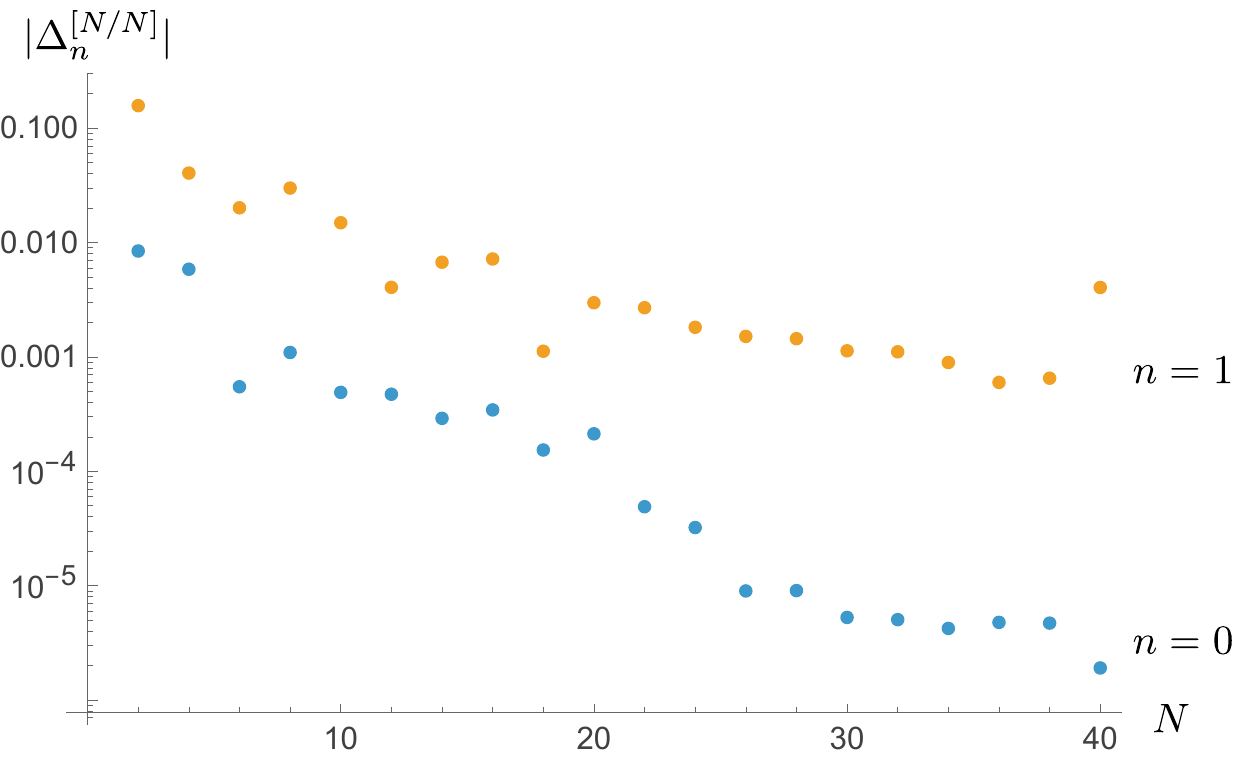}\\
    {\footnotesize (a) $(\ell,m)=(0,0)$.}
  \end{minipage}
    \begin{minipage}[b]{0.45\linewidth}
    \centering
    \includegraphics[width=0.95\linewidth]{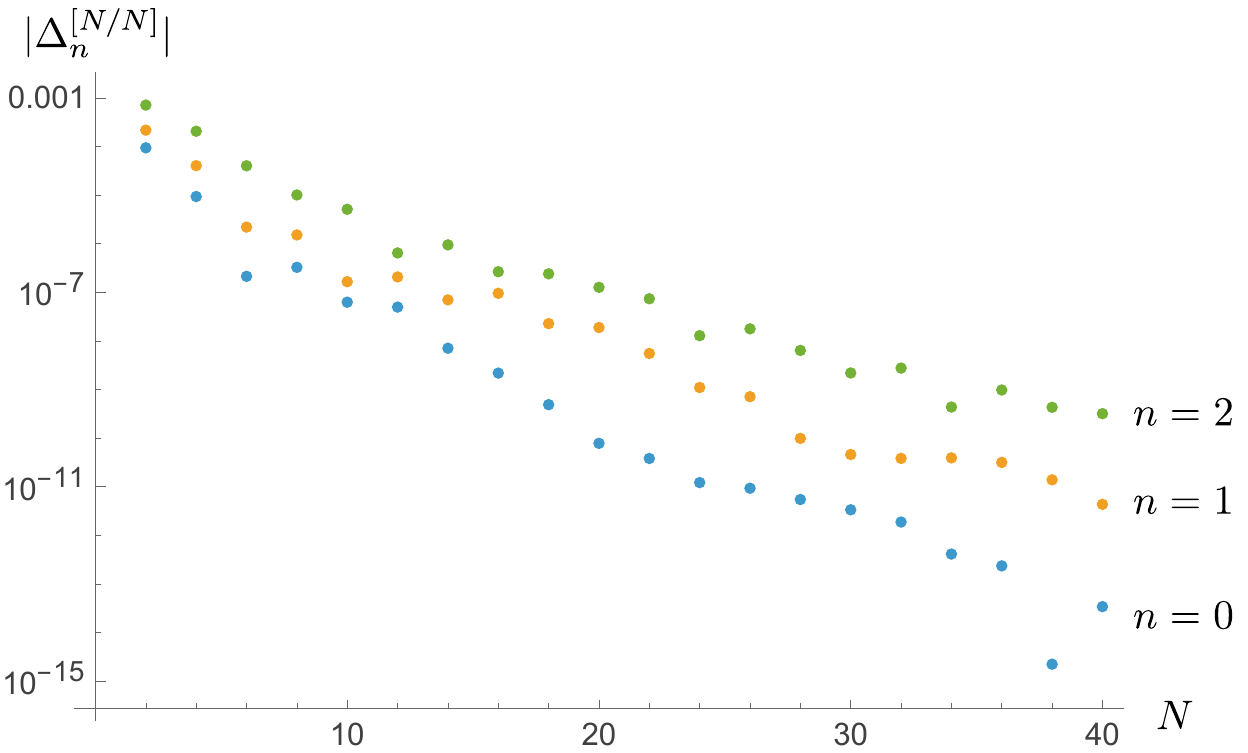}\\
    {\footnotesize (b) $(\ell,m)=(1,0)$.}
  \end{minipage} \\
    \begin{minipage}[b]{0.45\linewidth}
    \centering
    \includegraphics[width=0.95\linewidth]{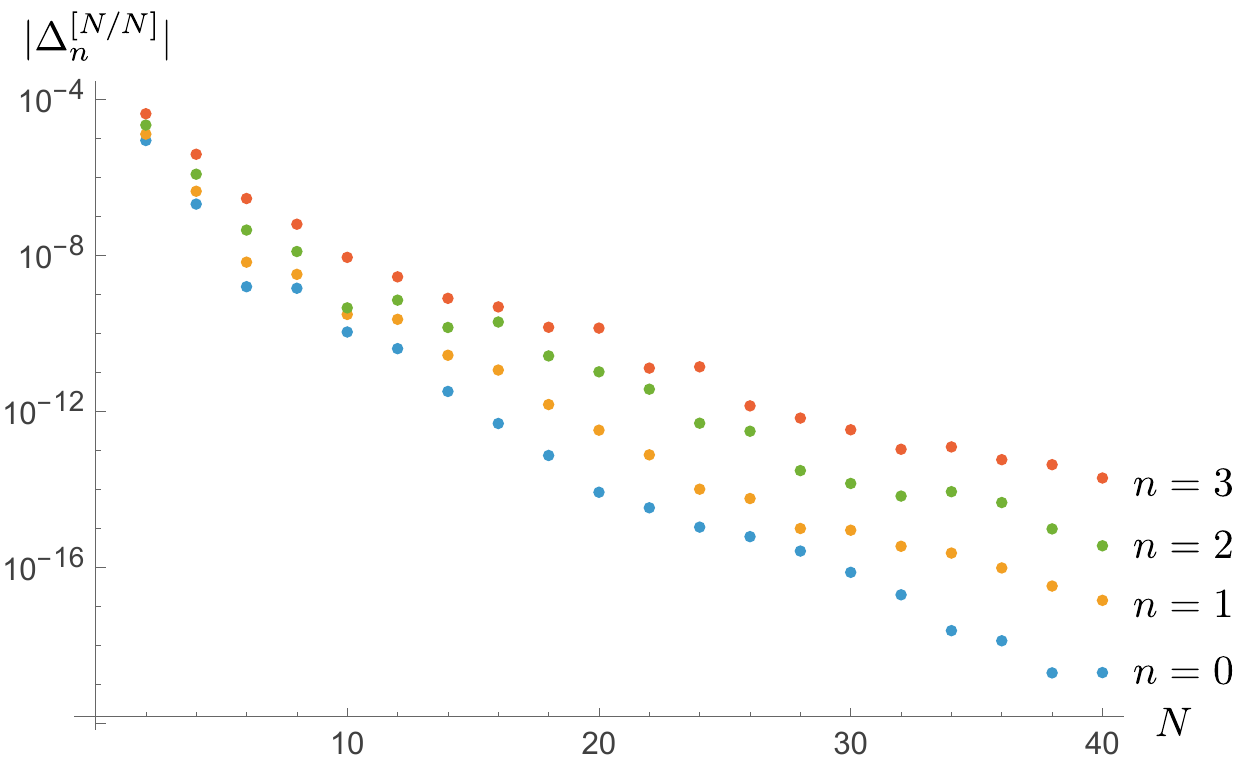}\\
    {\footnotesize (c) $(\ell,m)=(2,0)$.}
  \end{minipage}
  \begin{minipage}[b]{0.45\linewidth}
    \centering
    \includegraphics[width=0.95\linewidth]{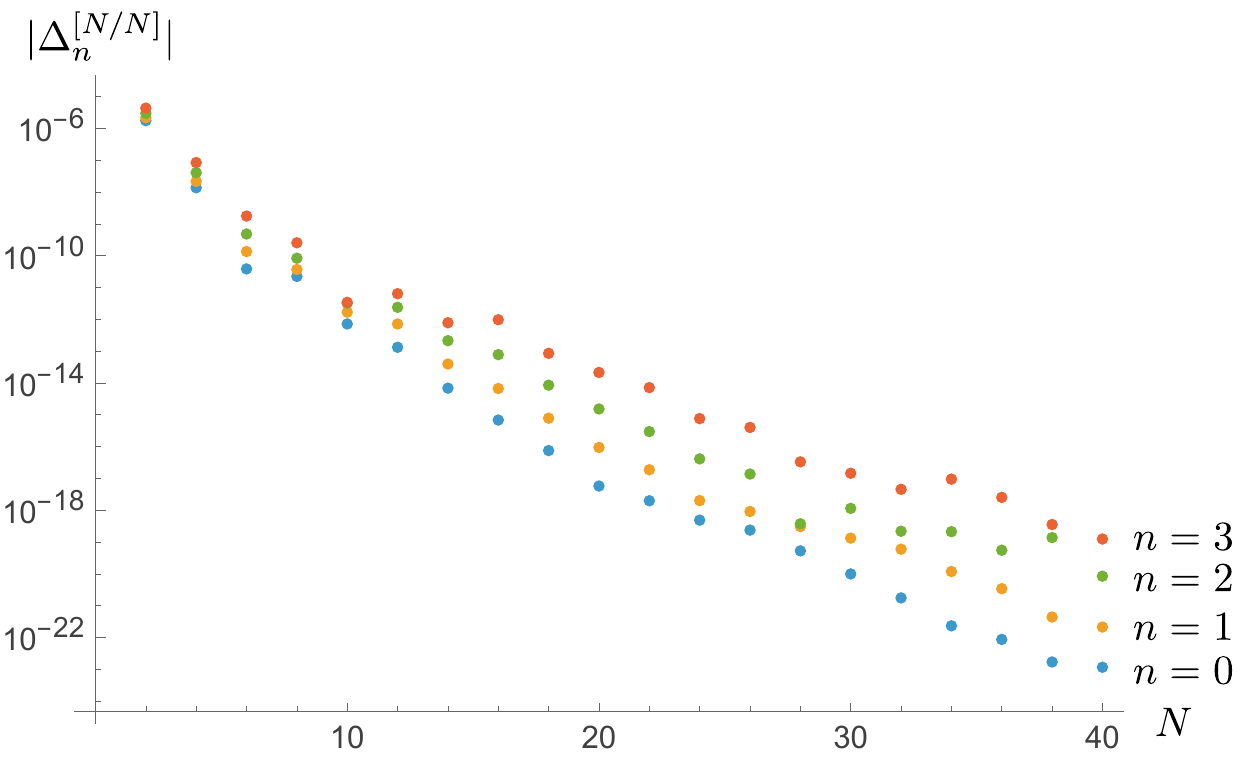}\\
    {\footnotesize (d) $(\ell,m)=(3,0)$.}
  \end{minipage}
  \caption{These graphs test the Borel--Pad\'e resummed quantization condition for the extremal Kerr black hole with $m=0$. We substitute highly accurate QNM frequencies into $\Delta_n^{[N/N]}$. The deviation of $\Delta_n^{[N/N]}$ from zero provides a measure of the accuracy of the quantization condition.}
  \label{fig:Delta-Kerr}
\end{figure}

Figure~\ref{fig:Delta-Kerr} tests the EQC in the axisymmetric sector $m=0$.
It shows the deviation of the diagonal Borel--Pad\'e resummation
$\Delta_n^{[N/N]}$ from zero for $\ell=0,1,2,3$.
In all cases, $\Delta_n^{[N/N]}$ decreases as the Pad\'e order $N$ is increased,
and the convergence becomes faster for larger $\ell$, as in the conventional
WKB analysis.

\begin{figure}[tb]
  \begin{minipage}[b]{0.45\linewidth}
    \centering
    \includegraphics[width=0.95\linewidth]{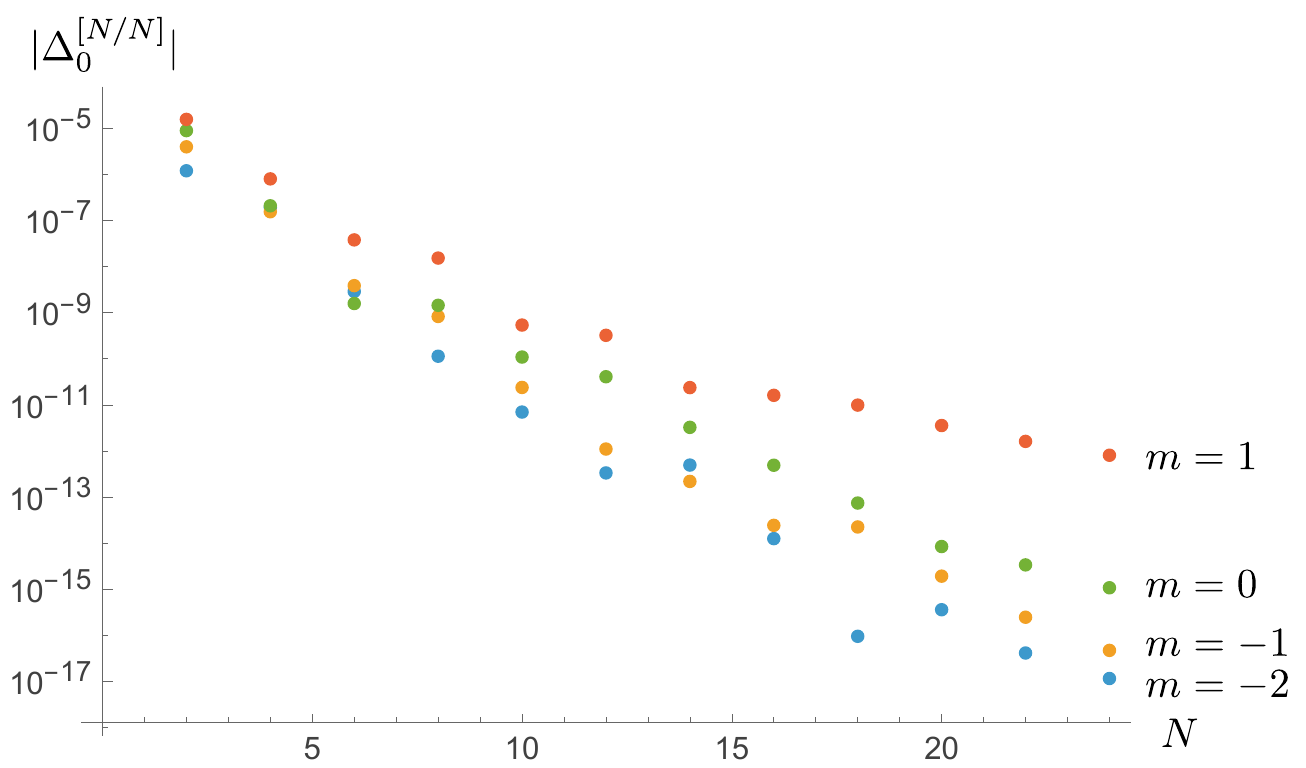}\\
    {\footnotesize (a) $(\ell,n)=(2,0)$.}
  \end{minipage}
    \begin{minipage}[b]{0.45\linewidth}
    \centering
    \includegraphics[width=0.95\linewidth]{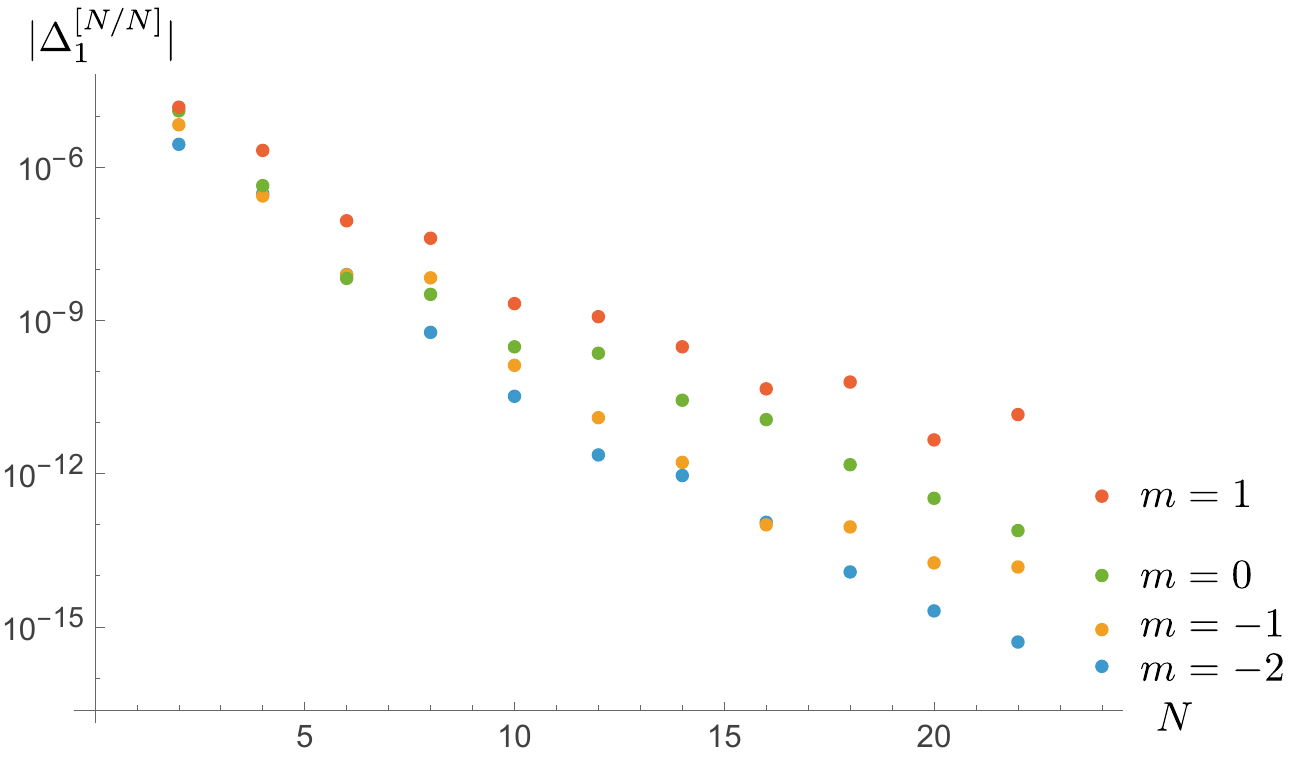}\\
    {\footnotesize (b) $(\ell,n)=(2,1)$.}
  \end{minipage}
  \caption{These graphs test the Borel--Pad\'e resummed quantization condition for the extremal Kerr black hole with $\ell=2$. We substitute highly accurate QNM frequencies into $\Delta_n^{[N/N]}$. The deviation of $\Delta_n^{[N/N]}$ from zero provides a measure of the accuracy of the quantization condition.}
  \label{fig:Delta-Kerr-m}
\end{figure}

We also examine the dependence on the azimuthal number $m$ by fixing
$\ell=2$.
As shown in Fig.~\ref{fig:Delta-Kerr-m}, the same decreasing behavior is
observed for $m=1,0,-1,-2$, both for the fundamental mode and for the first
overtone.
The exceptional case $m=\ell=2$ is excluded here because its Stokes geometry
is different from the non-exceptional cases discussed above.

\section{Conclusion}
In this paper, we applied exact WKB analysis to the QNM problem for scalar perturbations of four-dimensional asymptotically flat extremal black holes. For both the extremal Reissner--Nordstr\"om and Kerr backgrounds, the radial equations reduce to the doubly confluent Heun equation with two irregular singular points, and the QNM boundary conditions can be translated into exact quantization conditions. The central object in this formulation is the Borel resummed quantum period associated with the relevant cycle.

A main advantage of the extremal limit is that these quantum periods can be computed analytically to very high orders. In the extremal RN case, using the relation to the $N_f=2$ quantum Seiberg--Witten curve, we obtained the formal power series of the quantum period to very high order and performed its Borel--Pad\'e resummation. The resulting resummed quantization condition reproduces the QNM frequencies with high precision. We then carried out the analogous analysis for the extremal Kerr black hole. Owing to the similar Stokes geometry, the same exact quantization condition applies, while the quantum periods can be evaluated after an appropriate parameter identification. We again found good agreement between the resummed quantization condition and numerical QNM data.

Our results show that exact WKB provides a practical analytic framework for black hole QNMs, and that the direct evaluation of quantum periods is especially powerful when the spectral curve simplifies, as in the extremal case. There are several natural directions for future work. It would be interesting to extend the present analysis to non-extremal black holes, where the computation of the quantum periods is more involved, as well as to gravitational and electromagnetic perturbations. 
A preliminary result for the quantum period of the Schwarzschild black hole was obtained in \cite{Hatsuda:2021gtn}.
It is also desirable to clarify the TBA description of the resummed quantum periods in the present black hole setting. The TBA for the $SU(2)$ $N_f=2$ quantum Seiberg--Witten theory has been studied in~\cite{Imaizumi:2021cxf}, and this is expected to be useful in the present context (see also~\cite{Fioravanti:2021dce}). Another interesting problem is to analyze the behavior of QNMs at high overtone number from the EQC obtained in this paper. In the extremal Kerr case, the sector with $m=\ell>0$ appears to be subtle, and a more careful treatment of this case remains an important open problem.

The exact WKB approach developed here may also be applicable beyond the computation of QNMs, since the connection coefficients directly encode the scattering data~\cite{Bonelli:2021uvf}. In particular, it would be worthwhile to explore possible applications to greybody factors and, more ambitiously, to excitation factors~\cite{Berti:2006wq,Zhang:2013ksa} through the frequency dependence of the corresponding connection formulae. %Another natural direction is to compare the connection formulae in exact WKB with those for Heun-type equations proposed in~\cite{Bonelli:2022ten}.

\acknowledgments
We thank Masashi Kimura and Ryo Namba for useful discussions. 
We also acknowledge the use of ChatGPT (OpenAI) and Gemini (Google DeepMind), which were helpful for language editing, improving readability, informal discussions, and auxiliary checks. We critically reviewed all outputs and take full responsibility for the content of this manuscript. The work of Y. H. was supported in part by JSPS KAKENHI Grant Nos. 22K03641 and 23K25790. 

\appendix

\section{Analytic Computations of Quantum Periods}\label{app:ACQP}
In this appendix, we explain how to obtain the high-order quantum periods analytically.
We use the $N_f=2$ quantum SW curve. Throughout this section, we consider the equal mass case
\begin{align}
m_1=m_2=\mu.
\end{align}
The Schr\"odinger equation is given by
\begin{align}
\biggl(-\hbar^2 \frac{d^2}{dz^2}+Q_0(z)+\hbar^2 Q_2(z) \biggr) \psi(z)=0,
\label{eq:Sch-app}
\end{align}
where
\begin{align}
Q_0(z)=\frac{\Lambda^2}{z^4}+\frac{2\mu \Lambda }{z^3}-\frac{2u}{z^2}+\frac{2 \mu\Lambda}{z}+\Lambda^2,\qquad
Q_2(z)=-\frac{1}{4z^2}.
\label{eq:Q0Q2-app}
\end{align}
Sometimes, it is more convenient to use the original $x$-variable expression:
\begin{equation}
\begin{aligned}
\biggl( -\hbar^2 \frac{d^2}{dx^2}+2\Lambda^2 \cosh 2x+4\mu\Lambda \cosh x \biggr) \varphi(x)=2u \varphi(x).
\end{aligned}
\end{equation}
We emphasize that both expressions give the same quantum periods.
We further rewrite it as
\begin{align}
\biggl( -\hbar^2 \frac{d^2}{dx^2}+(2\Lambda \cosh x+\mu)^2 \biggr) \varphi(x)=E \varphi(x),
\end{align}
where
\begin{align}
E:=2u+2\Lambda^2+\mu^2=A^2.
\end{align}
By using the rescaling freedom \eqref{eq:rescale}, we set $\Lambda=1$.
For $0<\mu<A-2$, the four turning points of $Q_0(z)$,
\begin{align}
\alpha_\pm &=\frac{-A-\mu\pm \sqrt{(A+\mu-2)(A+\mu+2)}}{2} ,\\
\beta_\pm &=\frac{A-\mu \pm \sqrt{(A-\mu-2)(A-\mu+2)}}{2} ,
\end{align}
are all real. We have $\alpha_-<\alpha_+<0<\beta_-<\beta_+$.
Note that $\log \beta_\pm$ correspond to two real turning points in the $x$-variable system.
We compute the quantum periods in this situation, and then perform analytic continuation to complex parameters.

At the classical level, we would like to evaluate the integral:
\begin{align}
\Pi_B^{(0)}&:=i\oint_{\gamma_{\beta_- \beta_+}}\sqrt{Q_0(z)}dz=2\int_{\beta_-}^{\beta_+} \frac{\sqrt{(z-\alpha_-)(z-\alpha_+)(z-\beta_-)(\beta_+-z)}}{z^2} dz.
\end{align}
It is not easy to do so. For this purpose, we perform the following change of variables:
\begin{align}
s=\frac{z+z^{-1}}{2}.
\end{align}
Then, we obtain the relatively simple expression:
\begin{align}
\Pi_B^{(0)}=4\int_{1}^\beta \sqrt{\frac{E-(2s+\mu)^2}{s^2-1}}ds,
\end{align}
where we have used the symmetry $\beta_+\beta_-=1$ and
\begin{align}
\alpha=\frac{-A-\mu}{2},\qquad \beta=\frac{A-\mu}{2}.
\end{align}
We see that the derivative of $\Pi_B^{(0)}$ with respect to $E$ can be simply written in terms of the complete elliptic integral of the first kind:
\begin{align}
\frac{d}{dE}\Pi_B^{(0)}=4\int_1^\beta \frac{ds}{\sqrt{(s^2-1)(E-(2s+\mu)^2)}}
=\frac{8}{\sqrt{(-\alpha+1)(\beta+1)}}\eK(\mathsf{m}). \label{eq:dPiB0}
\end{align}
The classical period $\Pi_B^{(0)}$ itself does not have such a simple structure, but we can still write down its analytic form in terms of the elliptic integrals, as in \eqref{eq:PiB0}.
Once we obtain the exact results, we can easily analytically continue them to complex domains.
We can define other classical periods,
\begin{align}
\Pi_A^{(0)}&:=4\int_{-1}^1 \sqrt{\frac{E-(2s+\mu)^2}{1-s^2}} ds, \\
\Pi_{\widetilde{B}}^{(0)}&:=4 \int_{\alpha}^{-1} \sqrt{\frac{E-(2s+\mu)^2}{s^2-1}} ds,
\end{align}
where $\alpha<-1$ and $\beta>1$ for $0<\mu<A-2$. These are also exactly given by
\begin{align}
\Pi_A^{(0)}&=8\sqrt{(-\alpha+1)(\beta+1)}\biggl[ \eE(\mathsf{m'})+\biggl(1-\frac{\mathsf{m}'}{\mathsf{n}'}\biggr)\eK(\mathsf{m}')\notag \\
&\hspace{4truecm}-\biggl(2-\mathsf{n}'-\frac{\mathsf{m}'}{\mathsf{n}'} \biggr)\ePi(\mathsf{n'}|\mathsf{m'})\biggr],\\
\Pi_{\widetilde{B}}^{(0)}&=\Pi_B^{(0)}+4\pi \mu,
\end{align}
where
\begin{align}
\mathsf{m}'=\frac{2(\beta-\alpha)}{(-\alpha+1)(\beta+1)},\qquad \mathsf{n}'=\frac{\beta-\alpha}{\beta+1}.
\end{align}

It is well-known that the classical period integral $\Pi_\gamma^{(0)}$ for a given cycle $\gamma$ satisfies the so-called Picard--Fuchs equation. In the present case, the Picard--Fuchs equation is given by
\begin{align}
\biggl( P(E) \frac{d^2}{dE^2}+Q(E) \frac{d}{dE}+R(E) \biggr) \frac{d\Pi_\gamma^{(0)}}{dE}=0,
\end{align}
where
\begin{align}
P(E)&=4E(E+\mu^2-4)\bigl(E-(\mu-2)^2\bigr)\bigl(E-(\mu+2)^2\bigr), \\
Q(E)&=4\bigl( 2E^3+(\mu^2-20)E^2-4(\mu^4-16)E+(\mu^2-4)^3 \bigr),\\
R(E)&=E^2+(3\mu^2-8)E-2(\mu^2+2)(\mu^2-4).
\end{align}
The Picard--Fuchs equation is very useful to see the behavior of the period in various parameter regimes.

Let us now turn to the computation of the quantum periods. Here we employ an efficient method, known as the differential operator approach~\cite{Mironov:2009uv, Ito:2017iba}.
It is known that the quantum periods can be obtained by acting with differential operators on the classical period:
\begin{align}
\Pi_\gamma^{(k)}=\mathcal{D}^{(k)} \Pi_\gamma^{(0)}.
\end{align}
This relation is derived from the fact that at the integrand level, we generically have
\begin{align}
S_{2k-1}(z)=\mathcal{D}^{(k)} S_{-1}(z)+\frac{d}{dz} (\cdots),
\end{align}
where $\cdots$ denotes an algebraic function of $z$.
For $k=1$, we find
\begin{equation}
\begin{aligned}
\mathcal{D}^{(1)}=-\frac{2}{3}\del_E+\biggl( \frac{53+13\mu^2}{3}-\frac{79}{12}E \biggr)\del_E^2
+\biggl[ -2(4-\mu^2)^2+8(4+\mu^2)E-6E^2 \biggr]\del_E^3\\
+\biggl[ -(4-\mu^2)^2 E+2(4+\mu^2)E^2-E^3 \biggr] \del_E^4
\end{aligned}
\end{equation}
For $k \geq 2$, we observe the following simple structure:\footnote{Note that these differential operators are not fixed uniquely because the classical period satisfies the Picard--Fuchs equation. Here we look for the differential operators so that their coefficients are given by polynomials in $E$.}
\begin{align}
\mathcal{D}^{(2)}=\sum_{j=2}^5 d_j^{(2)}(E) \del_E^j,\qquad
\mathcal{D}^{(3)}=\sum_{j=4}^7 d_j^{(3)}(E) \del_E^j,\qquad
\mathcal{D}^{(k \geq 4)}=\sum_{j=k}^{2k} d_j^{(k)}(E) \del_E^j,
\end{align}
where $d_j^{(k)}(E)$ is a polynomial in $E$ of degree at most three.
If we assume this ansatz, then we can fix $\mathcal{D}^{(k)}$ up to high $k$ efficiently.
An important consistency check of this computation is that, for $\mu=0$, the quantum periods should reduce to the modified Mathieu case (i.e.\ the $N_f=0$ case). We have verified that this is indeed the case.

In the above computation, we restricted ourselves to ordinary differential operators in $E$. This is useful for our purpose, since the value of $\mu$ can be fixed from the outset. However, one can also construct partial differential operators involving both $E$ and $\mu$ derivatives. Such operators have a simpler structure~\cite{Mironov:2009uv,Ito:2017iba}. In the present case, we find
\begin{align}
\mathcal{D}^{(1)}%&=-\biggl( \frac{1}{6}u \del_u+\frac{1}{8} \mu \del_\mu+\frac{1}{12} \biggr) \del_u \\
&=-\frac{1}{24}( 4u \del_u+3 \mu \del_\mu+2) \del_u, \\
\mathcal{D}^{(2)}&=\frac{1}{5760}
( 112u^2\partial_u^2+480u\partial_u+168\mu u\partial_\mu\partial_u+63\mu^2\partial_\mu^2+345\mu\partial_\mu
+300)\partial_u^2, 
\end{align}
where we have used the original parameter $u$ instead of $E$, since the coefficients become much simpler in this parametrization.

\bibliographystyle{amsmod}
\bibliography{EWKB}

\end{document}